\documentclass[11pt]{article}
\pdfoutput=1

\usepackage{lineno,hyperref}
\usepackage{color}
\usepackage{amsfonts}
\usepackage{amsmath}
\usepackage{amssymb}
\usepackage{amsthm}
\usepackage{subeqnarray}
\usepackage{lscape,graphicx,url}
\usepackage{subcaption}
\usepackage{anysize}
\usepackage{proof}
\usepackage{fancyhdr}
\usepackage{latexsym}
\usepackage{array}
\usepackage{multirow}
\usepackage{textcomp}
\usepackage{optprog}
\usepackage{mathtools}
\usepackage{proof}
\usepackage{enumitem}
\usepackage{csquotes}
\usepackage{centernot}
\usepackage{bm}
\usepackage{array}
\usepackage{natbib}
\usepackage[margin=2.2cm]{geometry}
\newcolumntype{H}{>{\setbox0=\hbox\bgroup}c<{\egroup}@{}}

\renewcommand{\vec}[1] {\mbox{\boldmath$#1$}}

\newcommand{\N}{\overline{N}}

\hypersetup{
    colorlinks=true,    
    linkcolor=blue,     
    citecolor=blue,     
    urlcolor=blue,      
    pdfborder={0 0 0}   
}

\title{Portfolio optimisation: bridging the gap between theory and practice}

\author{Cristiano Arbex Valle$^1$}

\begin{document}

\maketitle

\begin{center} 
{\footnotesize

$^1$Departamento de Ci\^{e}ncia da Computa\c{c}\~{a}o, \\
 Universidade Federal de Minas Gerais, \\
 Belo Horizonte, MG 31270-010, Brasil \\
 arbex@dcc.ufmg.br
}
\end{center} 

\begin{abstract}

Portfolio optimisation is essential in quantitative investing, but its implementation faces several practical difficulties. One particular challenge is converting optimal portfolio weights into real-life trades in the presence of realistic features, such as transaction costs and integral lots. This is especially important in automated trading, where the entire process happens without human intervention.

Several works in literature have extended portfolio optimisation models to account for these features. In this paper, we highlight and illustrate difficulties faced when employing the existing literature in a practical setting, such as computational intractability, numerical imprecision and modelling trade-offs. We then propose a two-stage framework as an alternative approach to address this issue. Its goal is to optimise portfolio weights in the first stage and to generate realistic trades in the second. Through extensive computational experiments, we show that our approach not only mitigates the difficulties discussed above but also can be successfully employed in a realistic scenario.

By splitting the problem in two, we are able to incorporate new features without adding too much complexity to any single model. With this in mind we model two novel features that are critical to many investment strategies: first, we integrate two classes of assets, futures contracts and equities, into a single framework, with an example illustrating how this can help portfolio managers in enhancing investment strategies. Second, we account for borrowing costs in short positions, which have so far been neglected in literature but which significantly impact profits in long/short strategies. Even with these new features, our two-stage approach still effectively converts optimal portfolios into actionable trades.

\end{abstract}

{\bf Keywords:} portfolio optimisation, exogenous constraints, transaction lots, futures contracts, quantitative finance

\section{Introduction}

Portfolio optimisation is widely acknowledged for its significance in investment decision-making. Its goal is intuitive and conceptually simple: deciding the proportion of capital to be invested in each available asset. In practice, for its successful use, it is necessary to convert its output (suggested proportions) into a list of trades required to purchase the desired portfolio.

At first glance, this problem may seem somewhat trivial, but the complexities of real-life trading mean that naive solutions, such as simple rounding, may lead to portfolios that significantly diverge from the optimal allocation. This is especially important when fund management is entirely automated, with decisions controlled by algorithms. Here it is essential for the optimisation to be integrated with the actual purchases and sales of assets to ensure alignment with the desired portfolio structure.

Overall, practitioners of portfolio optimisation must consider complicating factors such as (i) regulatory and internal investment policies, (ii) costs that directly affect investment returns and (iii) specific rules for trading different asset types.  As an example, buying a stock confers partial ownership of a company, whereas buying derivatives requires only signing a contract and maintaining a margin account as collateral. Additionally, some assets may not be easily liquidated if not traded in integral lots. Other assets, such as fixed income instruments far from maturity, may incur significant losses if liquidated prematurely.

In the scientific literature, several of the issues above have been addressed with exogenous (real-world) constraints, incorporated as extensions of portfolio optimisation models. Among these are exposure limits, restrictions on portfolio turnover, lots and various transaction cost functions. Other issues have, to the best of our knowledge, not been addressed explicitly, such as the specificities of futures contracts or fixed income instruments in portfolio optimisation. Nevertheless, there is currently an established and well-known body of constraints that deal with many problems of practical investing. For detailed reviews, we refer the reader to \cite{mansini2014, mansini2015, kolm2014}.

Most formulations of portfolio selection problems are relative, i.e. they employ portfolio weights as decision variables and ignore the value of the capital available for investment. Certain real-world features (such as costs and lots), however, are often more aptly described with financial values instead. These may require absolute formulations, where decision variables are portfolio holdings - the number of shares to be held in each asset. A mixed formulation uses both sets of variables.

With the explicit intention of converting portfolio weights into feasible holdings, in this paper we focus on features that are best represented with absolute or mixed formulations. We examine how the current literature approaches this problem, with an empirical example illustrating the practical difficulties of applying them in practice. To address these issues, we propose an alternative two-stage approach: in the first stage, we optimise portfolio weights, and in the second, we determine optimal holdings. Breaking up the problem in two effectively mitigates the issues seen in prior work and allows for a wider range of real-world features to be included without adding too much complexity to any single model. With this added flexibility, we model two novel real-world features: borrowing costs in short positions and the construction of portfolios that include both equities (such as stocks and exchange-traded funds) and futures contracts.

Borrowing costs have been mostly overlooked when discussing long/short portfolio models. In practice, however, these costs can significantly impact profits in long/short strategies. While theoretically they can be modelled as an extension of single-stage models from the existing literature, solving these models - even without borrowing costs - is already often impractical, as we show empirically in this paper.

Futures contracts (futures for short) comprise some of the most liquid financial instruments worldwide. Their diverse nature gives portfolio managers better options to increase diversification. They can be easily acquired in short or leveraged positions, offering useful tools for investment decision making. In this paper, we highlight their operational differences and adapt our two-stage framework to account for them, illustrating their potential benefits with empirical examples.

The two stages are solved sequentially: the first stage formulation deals with equities and futures and supports several applicable real-world constraints. In the second stage, we employ a goal-programming approach to minimise the deviation between the actual portfolio holdings and the optimal weights determined in the first stage. We consider lots, costs and specific features of futures contracts, such as leverage, to produce feasible portfolio holdings. Additionally, we propose a way to balance the conflicting objectives of minimising deviation and transaction costs within a single objective function.

We present extensive computational experiments to validate our framework. We compare our two-stage approach with a mixed formulation from the literature \citep{woodside2013}. For reasonably-sized portfolios, we evaluate (i) how closely the second stage approximates the solutions from the first stage, (ii) the computational effort required and (iii) how it handles the trade-off between minimising deviation and transaction costs spent. 

Our key contributions are summarised below:

\begin{itemize}
\item We identify and discuss the limitations of existing models, demonstrating empirically their impracticality for real-world application,
\item We propose a two-stage framework that effectively addresses these limitations,
\item We design the framework to explicitly account for equities and futures contracts, and discuss why it may be advantageous to do so,
\item We incorporate borrowing costs in short positions during the second stage,
\item We empirically show that, despite the inclusion of these new features, the problem remains computationally tractable.
\end{itemize}

\noindent An added benefit of the computational tractability of our approach is that it is possible to simulate strategies more faithfully, more accurately reflecting what would have happened if real capital had been allocated to a given financial strategy.

The remainder of this paper is organised as follows. In Section \ref{sec:literature}, we present a review of the relevant literature. In Section \ref{sec:issues} we discuss the limitations of existing models. In Section \ref{sec:futures}, we discuss aspects of futures contracts that differentiate them from equities and provide an empirical example demonstrating how integrating these asset classes into a single portfolio can benefit portfolio managers. In Sections \ref{sec:firstStage} and \ref{sec:secondStage}, we present our framework, with the former introducing the first stage and the latter introducing the second stage. We extend our framework to account for borrowing costs in Section \ref{sec:borrowingCosts}. In Section \ref{sec:experiments} we empirically evaluate our second stage and in Section \ref{sec:managerial} we discuss some of its limitations. Finally, in Section \ref{sec:conclusions} we present our concluding remarks.

\section{Literature review}
\label{sec:literature}

Consider a fund that already possesses a certain portfolio, whose value is given by the amount held in cash plus the amount held in each asset at their current market price. In a rebalancing (revision) day, the optimisation model suggests new portfolio proportions. The managers have to buy/sell some assets in order to update the portfolio, for which they will pay the applicable transaction costs. The costs are deducted directly from the portfolio itself, and the remaining value becomes the reference point for the new optimised proportions. Because many assets are traded in lots, the exact desired portfolio proportions are usually not attainable, so there must be a policy to account for this feature: either during the optimisation (the solution adopted by the existing literature) or afterwards (as we propose here in this paper).

In this section, we provide a brief, non-exhaustive overview of the literature on exogenous constraints, with a particular focus on those related to lots and transaction costs. For more comprehensive reviews we refer the reader to \cite{mansini2014, mansini2015, kolm2014}. In Section \ref{sec:issues}, we discuss the rationale behind adopting an alternative approach for dealing with these features.

Exogenous constraints were first addressed by \cite{pogue1970}, who considered short sales, transaction costs, liquidity and taxes in the context of Modern Portfolio Theory (MPT) \citep{markowitz1952}. Turnover constraints, which limit how much a portfolio can change during a rebalance, were first introduced by \cite{schreiner1980}. Another early relevant work was that of \cite{rudd1979}.

\cite{speranza1996} proposed a linear-programming formulation for finding the portfolio that minimises the semi-$L_1$ dispersion measure, subject to several classes of exogenous constraints. The authors included lots, fixed and variable transaction costs, upper exposure bounds on individual assets and portfolio cardinality. A novelty of their formulation was the use of portfolio holdings as decision variables, which allowed the authors to account for optimal proportions with respect to the discounted portfolio value.

\cite{mansini1999} proposed a similar model, however transaction costs were discounted from assets expected returns rather than the portfolio value, with the goal of preventing the underestimation of risk or the overestimation of return. This approach complicates the generation of feasible trades as proportions are based on the undiscounted portfolio value. In related models, \cite{li2006} deduct costs directly from asset returns and \cite{kellerer2000} adds fixed transaction costs as a penalty in the expected return constraint. \cite{chiodi2003}, \cite{mansini2005} and \cite{angelelli2008} also deducted costs from expected returns, however they considered extensions such as including a capital-gains tax. \cite{guastaroba2009} considered portfolio rebalancing and transaction costs, but did not consider lots.

\cite{bertsimas1999} proposed a two-stage approach where portfolio weights are chosen first and then a Mixed-Integer Programming (MIP) problem is solved to approximate the target portfolio. The authors combined many goals into a single objective function, including return, transaction costs, liquidity, number of transactions and closeness to the target portfolio. They also consider rebalances, where an existing portfolio is currently held and must be converted into a new portfolio. They adopt a somewhat similar two-stage approach to the one we propose here, but with key differences: their second stage also employs proportions as decision variables and it combines objectives that influence portfolio weights, which we consider as first stage decisions. Due to this, their proposal does not directly fit the idea of generating tradable portfolios. We believe that the two approaches are not directly comparable, but we discuss it further in Section \ref{sec:managerial}.

\cite{chang2000} developed heuristics for the cardinality-constrained portfolio selection problem, with upper exposure bounds, class constraints and buy-in thresholds. Class constraints enforce minimum and maximum exposure levels in subsets of assets. Buy-in thresholds help prevent (very) small positions by ensuring minimum holding levels. \cite{jobst2001} and \cite{mitra2003} considered lots in a relative model, combining the works of \cite{chang2000} and \cite{mansini1999}.

\cite{jacobs2005} proposed separating decision variables into two sets - one for long and one short positions - in order to impose shorting limits. \cite{kumar2010} proposed a mixed formulation with long-short separation to solve the Markowitz quadratic model with the presence of shorting limits and risk-free lending and borrowing. \cite{valle2014b} considered constraints on Regulation T, which prevents the proceedings of short sales from being used to purchase other long positions. In this paper we adopt a similar first stage approach as \citeauthor{jacobs2005} and \citeauthor{valle2014b} in order to model futures contracts.

\cite{lin2008} proposed a non-convex mixed formulation that addresses lots, but not transaction costs. \cite{bonami2009} extended their work by introducing a cash asset representing uninvested capital, but also without costs. Similarly, \cite{bartholomew2009} introduced a non-convex mixed formulation in the context of MPT, and which includes buy-in thresholds and lots, but no transaction costs.

\cite{woodside2013} proposed a mixed model with transaction costs, buy-in thresholds, exposure bounds, cardinality constraints and portfolio rebalancing. The model can be easily extended to incorporate lots. \cite{valle2014a} adopted a similar approach and raised the matter of unavoidable transaction costs, when assets held in the current portfolio must be liquidated. In their work transaction costs are explicitly minimised in a later stage.

\section{Difficulties with the existing literature}
\label{sec:issues}

Let $N$ be the set of assets in which we may invest, and let $w_i$ be the (unknown) proportion to be invested in asset $i \in N$. Suppose that the return of asset $i$ is modelled with a discrete distribution with $S$ scenarios (realisations), where $r_{is}$ is the return on the $s^{\text{th}}$ scenario. For $s$, the weighted sum $\sum_{i \in N} w_i r_{is}$ gives the portfolio return for scenario $s$. When generalised to $S$ we have a discrete representation of the portfolio distribution. Most portfolio models optimise a metric of said distribution: the standard Markowitz model, for instance, minimises $\vec{w}^T \Sigma \vec{w}$ where $\vec{w} = (w_1, \dots, w_N)$ and $\Sigma$ is the covariance matrix.

Now consider an absolute formulation where $x_i$ is the number of lots/shares of $i$ to be held in the portfolio. These variables may be integer (if the asset requires lots) or fractional (if not, such as cryptocurrencies). Let $V_i$ be the current price of $i$. Such variables are ideal for representing features that require financial values, but modelling the statistics of the portfolio distribution is not always straightforward: portfolio variance, for instance, would be modelled as the difficult non-linear expression $\frac{\vec{x}^T \Sigma \vec{x}}{\left(\vec{x}^T \vec{V}\right)^2}$, where $\vec{x}^T \vec{V}$ is the portfolio value. Not every portfolio utility function can be trivially represented with an absolute formulation. We argue in this section that in order to optimise some generic portfolio model {\bf and} produce financial holdings as a result, neither a relative nor an absolute formulation suffices. 

Let $X_i$ be the current holdings in asset $i$ and $C$ be the total amount of funds to be added/withdrawn from the portfolio, with the total portfolio value prior to trading defined as $P = \sum_{i \in N} V_i X_i + C$. Let $f_i$ be the fractional costs associated with trading asset $i$, paid proportionally to the size of the trade. In a mixed formulation, the constraints that link $x_i$ and $w_i$ are:
{\small
\begin{align}
w_i & = \frac{x_i}{P - \sum_{i \in N} |x_i - X_i|f_i V_i} & \forall \; i \in N
\label{eqissues1}
\end{align}
}

\noindent In Equation \eqref{eqissues1}, $P - \sum_{i \in N} |x_i - X_i|f_i V_i$ represents the discounted portfolio value after the (unknown) trades take place. Due to its non-convexity, Equation \eqref{eqissues1} results in a computationally difficult mixed formulation. To make matters worse, Equation \eqref{eqissues1} could be extended with other costs functions such as fixed, piecewise-linear and borrowing costs. Additionaly, When $\vec{x}$ is (partially) integer, the budget constraint $\sum_{i \in N} w_i = 1$ is unlikely to be satisfied by a lot-feasible portfolio. In fact, \cite{angelelli2008} showed that optimal solutions to problems with and without lots may include different assets.

In order to prevent non-linearity, \cite{speranza1996} proposed an original absolute formulation with lots and transaction costs. Instead of variance, the authors showed that minimising the Mean Absolute Deviation \citep{konno1991} can be linearly solved with portfolio holdings instead of proportions. Let $g_i$ be the fixed costs associated with trading asset $i$, and let $z_i$ be a binary decision variable indicating whether any amount of $i$ is to be purchased in the portfolio. \cite{speranza1996} proposed the following budget constraint:
{\small
\begin{align}
P_0 \leq \sum_{i \in N} (1 + f_i) V_i x_i + g_i z_i & \leq P_1
\label{eqissues2}
\end{align}
}

\noindent where instead of equality, they define $P_0 \leq P \leq P_1$ as arbitrary lower and upper bounds on the value of the portfolio after trading. In particular, the presence of the lower bound $C_0$ can be conflicting with the main objective. In fact, \enquote{without the lower bound the optimisation of a risk function tends to invest the minimum amount of capital that guarantees the required expected return of the investment.} \citep{mansini2015}. Note also that the Equation \eqref{eqissues2} does not account for current holdings $X_i$.

\cite{mansini1999} showed that if $P_0 \neq P_1$, the portfolio selection problem is NP-Hard. Due to these computational difficulties, their formulation was simplified to prevent transaction costs from being explicitly discounted from $P$, with them rather being represented as discounts in the assets expected returns. \cite{kellerer2000} showed that an absolute formulation with variable and fixed costs, modelled with Equation \eqref{eqissues2}, is NP-hard even if no lots are considered (if $x_i$ can be fractional). New features were proposed by \cite{mansini2005}, but under similar modelling choices.

\cite{li2006} considered explicit costs and proposed solving the mean-variance problem with a difficult non-linear model. The authors developed a specialised algorithm, but which cannot be easily extended to a generic portfolio problem. \cite{angelelli2008} proposed minimising the Conditional Value-at-Risk (CVaR) with an absolute formulation, but did not explicitly consider costs. Computational difficulties motivated the authors to later develop a specialised heuristic \citep{angelelli2012}.

\cite{jobst2001} and \cite{mitra2003} proposed a relative formulation for modelling lots, but without considering transaction costs. Let $l_i$ be the fraction of $P$ that represents a lot of asset $i$. With $\gamma_i \in \mathbb{Z}$ as the number of lots to be held, the authors modelled the budget constraint as:
{\small
\begin{align}
\sum_{i \in N} l_i \gamma_i + d^- - d^+ & = 1 
\label{eqissues3}
\end{align}
}

\noindent where $d^-$ and $d^+$ are downward and upward deviations that are penalised in the objective function. Such penalty may be conflicting with other optimisation criteria. Also, as transaction costs are not considered, the estimation of $l_i$ is based on an arbitrary value as the total portfolio value {\bf after} trading is unknown. \cite{baumann2013} also proposed relative formulations with different objectives. Unlike \cite{jobst2001}, instead of deviations the authors added arbitrary bounds for linearising the budget constraint. They also discounted costs from the objective function, but not from the budget constraint.

It may be possible to model most (if not all) real-world features using relative formulations, provided that all financial values are represented as proportions of a reference portfolio value. However, accounting for the costs discount also results in non-linear expression. Even without considering it, the referred authors reported computationally intractable models which may also, under the common scenario where the sums involved are large, be prone to losses in numerical precision.

The remaining single stage alternative relies on employing mixed formulations. \cite{lin2008} modelled the budget constraint as $\sum_{i \in N} V_i x_i \leq P$ (without transaction costs) and the linking constraints as:
{\small
\begin{align}
w_i & = \frac{V_i x_i}{\sum_{i \in N} V_i x_i} & \forall i \in N \label{eqissues4}
\end{align}
}

\noindent which, even without costs, produces a difficult nonlinear optimisation model. Due to the integral constraints, part of $P$ may be left uninvested, but this amount is not considered in the formulation.

\cite{bonami2009} introduced a cash asset $w_c$, representing uninvested capital. With this variable, the authors linearised Constraints \eqref{eqissues4} with:
{\small
\begin{align}
w_i & = \frac{V_i x_i}{P} \label{eqissues5} \\
w_c + \sum_{i \in N} w_i & =  1 \label{eqissues6}
\end{align}
}

\noindent However, had transaction costs been included, the discounted $P$ would be unknown, rendering the problem non-linear. By omitting transaction costs, the optimal solution could yield a portfolio $(x_1, \dots, x_N)$ whose cumulative financial value is greater than the $P$ minus the discounts.

\cite{bartholomew2009} proposed a mixed non-convex formulation for the mean-variance problem with buy-in thresholds and lots, but without costs. While a specialised algorithm was developed, it is unclear whether it can be generalised to any optimisation criteria.

\cite{woodside2013} proposed a mixed formulation that  takes into account fixed and variable costs as well as current portfolio holdings, and which can be easily extended to include lots. Let $f^b_i$ and $g^b_i$ be the fractional and fixed costs associated with {\bf buying} shares of $i$, and, accordingly, let $f^s_i$ and $g^s_i$ be the costs associated with {\bf selling} shares of $i$. The authors define $f^b_i$ and $f^s_i$ as financial values rather than proportions. Let $y^b_i$ and $y^s_i$ be decision variables corresponding to the number of units of $i$ bought and sold respectively, and let $\alpha^b_i$ and $\alpha^s_i$ indicate whether any amount of $i$ is bought/sold. Variables $\alpha^b_i$ and $y^b_i$ as well as $\alpha^s_i$ and $y^s_i$ are linked through lower and upper bounds imposed on the amount of shares that can be sold/purchased.

Let $\text{TC} = \sum_{i \in N} \left( f^b_i y^b_i + f^s_i y^s_i + g^b_i \alpha^b_i + g^s_i \alpha^s_i \right)$ be the total transaction costs spent in a rebalance. The constraints ensuring rebalance consistency and costs discount are as follows:
{\small
\begin{align}
\alpha^b_i + \alpha^s_i & \leq 1 \label{eqissues7}\\
X_i + y^b_i - y^s_i & = x_i   & \forall i \in N \label{eqissues8}\\
\sum_{i \in N} V_i X_i + C & = \sum_{i \in N} V_i x_i + \text{TC} \label{eqissues9}\\
w_i & = \frac{V_i x_i}{\sum_{k \in N} V_i X_i + C}  & \forall i \in N \label{eqissues10} \\
w^{\text{tc}} & = \frac{\text{TC}}{\sum_{k \in N} V_i X_i + C} \label{eqissues11}\\
\sum_{i \in N} w_i + w^{\text{tc}} & = 1 \label{eqissues12} 
\end{align}
}

In Equation \eqref{eqissues11}, a weight $w^{\text{tc}}$ is assigned to the proportion of the original portfolio value spent in transaction costs. With this variable, Equation \eqref{eqissues10} remains linear, but the budget constraint \eqref{eqissues12} requires the addition of $w^{\text{tc}}$. Constraints \eqref{eqissues7}-\eqref{eqissues9} ensure consistency between the portfolio value before and after trading. Incorporating lots can be achieved by defining $x_i$ as integer variables.

A potential drawback with this approach is an implicit incentive to increase $w^{\text{tc}}$ and reduce $\sum_{i \in N} w_i$ as an artificial way to reduce risk. To prevent this, the authors impose an arbitrary upper limit on transaction costs spent, which may possibly underestimate the true costs associated with the risk-optimal portfolio of choice. This issue is similar to the one faced by \cite{speranza1996} when defining lower bound $P_0$. 

Nevertheless, the \cite{woodside2013} approach allows us to reasonably produce lot-feasible solutions in a single stage, with optimal proportions consistent with the discounted portfolio. There are two ``hidden'' issues, however, that appear when we attempt to solve this formulation in practice: its computational difficulty and its susceptibility to numerical inaccuracies. When the sums involved are large, the formulation combines coefficients of very different orders of magnitude, leading to imprecisions in floating-point calculations. We demonstrate this empirically in the next section.

\subsection{Empirical example}
\label{sec:woodsideIssue}

Consider the problem of selecting a portfolio that maximizes expected return, while ensuring that no more than 5\% of the total investment is allocated to any single asset. Its lot-unrestricted optimal solution can be trivially found by sorting the assets by expected return and allocating 5\% to each one of the top 20 assets. To solve the lot-constrained version of this problem, we use the \cite{woodside2013} model (hereafter referred to as Woodside)\footnote{The complete Woodside formulation for this problem can be found in the appendix available at \url{https://github.com/cristianoarbex/portfolioSimulationData}.}. As long as there are at least 20 assets with positive expected return, Woodside has no incentive to artificially increase $w^{\text{tc}}$.

Here and in other experiments, we employ a single dataset with S\&P500 stocks, taking into account survivorship bias - at any given rebalance, the only stocks available for investment are the index constituents at the time. We assume that S\&P500 stocks are either traded in round lots (consisting of 100 shares each) or in odd lots (any integral number of shares). With the increase in electronic trading, trading in odd lots in the U.S. is becoming easier due to increased liquidity\footnote{\url{https://www.investopedia.com/terms/r/roundlot.asp}, last visited $17^{\text{th}}$ September 2024.}. In other stock markets (such as Bovespa in Brazil or BSE in India) trading in odd lots is either not feasible or often illiquid.

Portfolios are rebalanced at the last trading day of each month and evaluated daily until the data is exhausted. The out-of-sample period spans from $25^{\text{th}}$ December 2012 up to $29^{\text{th}}$ December 2023. During this period, a total of 132 rebalances are calculated. We simulate an initial investment of $\$1\text{,}000\text{,}000$ at the end of 2012 and assume no cash additions/withdrawals throughout the investment lifespan. For each rebalance, we take the average of the previous 200 historical daily returns as the expected returns. We assume trading at the end of the day based on closing prices, 5 basis points as variable transaction costs and no fixed costs. The return from the end of the previous day to the end of the current day is included in the in-sample historical data. We impose a time limit of 15 minutes (900s) for each Woodside rebalance.

For all experiments here and elsewhere, we employ \cite{cplex} with default configurations as MIP solver. Our simulation tool is developed in Python and all optimisation models are developed in C++. We run all experiments in an Intel(R) Core(TM) i7-3770 CPU @ 3.90GHz with 8 cores, 8GB RAM and with Ubuntu 22.04.3 LTS as the operational system.

Let $\mu_i$ be the expected return of asset $i$. We define the objective function as $\max \left( \Theta \sum_{i \in N} \mu_i w_i \right)$, where $\Theta$ is a constant. The role of $\Theta$ is to simply scale the objective function in order to reduce the numerical errors faced by CPLEX when solving the problem. Results are shown in Table \ref{table1}.

\begin{table}[!ht]
\centering
{\scriptsize
\renewcommand{\tabcolsep}{1mm} 
\renewcommand{\arraystretch}{1.4} 
\begin{tabular}{|l|l|l|r|rrrrrr|rrrrrr|}
\hline
\multicolumn{1}{|c|}{\multirow[c]{2}{*}{Strategy}} & \multicolumn{1}{c|}{\multirow[c]{2}{*}{Lots}} & \multicolumn{1}{c|}{\multirow[c]{2}{*}{$\Theta$}} & \multicolumn{1}{c|}{\multirow[c]{2}{*}{\begin{tabular}{c}Sol.\\found\end{tabular}}} & \multicolumn{6}{c|}{Exp. return relative decline (\%)} & \multicolumn{6}{c|}{Computational time (s)}\\
\cline{5-16}
 &  &  &  & \multicolumn{1}{c}{Min} & \multicolumn{1}{c}{$P_{25}$} & \multicolumn{1}{c}{Avg} & \multicolumn{1}{c}{Median} & \multicolumn{1}{c}{$P_{75}$} & \multicolumn{1}{c|}{Max} & \multicolumn{1}{c}{Min} & \multicolumn{1}{c}{Avg} & \multicolumn{1}{c}{Median} & \multicolumn{1}{c}{$P_{90}$} & \multicolumn{1}{c}{Max} & \multicolumn{1}{c|}{TL}\\
\hline
\multicolumn{1}{|l|}{\multirow[c]{10}{*}{Long only}} & \multicolumn{1}{l|}{\multirow[c]{5}{*}{Odd}} & $1 \times 10^{0}$ &    132 &  28.58 &  86.20 &  88.74 &  90.32 &  93.04 &  98.43 &    0.1 &    0.1 &    0.1 &    0.1 &    0.4 &      0\\
 &  & $1 \times 10^{1}$ &    132 &  69.94 &  91.66 &  92.96 &  93.68 &  95.61 &  99.50 &    0.1 &    0.2 &    0.1 &    0.4 &    2.2 &      0\\
 &  & $1 \times 10^{2}$ &    132 &  93.55 &  98.15 &  98.58 &  98.94 &  99.46 &  99.98 &    0.1 &    0.2 &    0.1 &    0.3 &    3.5 &      0\\
 &  & $1 \times 10^{3}$ &    132 &  99.66 &  99.94 &  99.94 &  99.96 &  99.97 & 100.00 &    0.1 &    0.5 &    0.2 &    1.1 &    4.8 &      0\\
 &  & $1 \times 10^{4}$ &    132 &  99.93 &  99.96 &  99.97 &  99.97 &  99.97 & 100.00 &    0.1 &    3.3 &    0.7 &    2.7 & TL     &      1\\
\cline{2-16}
 & \multicolumn{1}{l|}{\multirow[c]{5}{*}{Round}} & $1 \times 10^{0}$ &    132 &  91.91 &  97.32 &  97.68 &  97.95 &  98.57 &  99.62 &    0.1 &    0.3 &    0.1 &    0.6 &    2.4 &      0\\
 &  & $1 \times 10^{1}$ &    131 &  92.50 &  98.65 &  98.79 &  99.00 &  99.21 &  99.61 &    0.1 &    7.4 &    0.2 &    1.5 & TL     &      3\\
 &  & $1 \times 10^{2}$ &    132 &  92.52 &  98.67 &  98.82 &  99.01 &  99.25 &  99.68 &    0.1 &   10.0 &    0.3 &    2.5 & TL     &      4\\
 &  & $1 \times 10^{3}$ &    130 &  92.52 &  98.67 &  98.81 &  99.01 &  99.24 &  99.68 &    0.1 &   12.1 &    0.4 &    1.8 & TL     &      5\\
 &  & $1 \times 10^{4}$ &    131 &  92.52 &  98.67 &  98.82 &  99.01 &  99.25 &  99.68 &    0.1 &    9.9 &    0.4 &    1.7 & TL     &      4\\
\hline
\end{tabular}
}
\caption{Woodside model results. Decline in expected returns (when compared to lot-unconstrained first stage solutions) and computational time  (with $P_{\alpha}$ as the $\alpha^{\text{th}}$ percentile).}
\label{table1}
\end{table}

Consider the first row of data in the table, with assets traded in odd lots. Since $\Theta = 1$, the model uses the original unscaled objective function. Under column {\bf Sol. found} we show how many rebalances produced a valid solution (in this case, all of them). Woodside optimal expected returns from each successful rebalance are divided by the corresponding lot-unconstrained solutions in order to produce the statistics available under label {\bf Exp. return relative decline (\%)}. In the columns labeled {\bf Computational Time}, we present statistics on the time required to solve the Woodside model. The number of rebalances without proven optimality after 900s is shown in column {\bf TL} (Time Limit) - in the first row, all rebalances were optimally solved within 0.4s.

In that same row, a particular rebalance produced a Woodside optimal portfolio whose expected return was only 28.58\% of the corresponding lot-unconstrained optimal expected return. On average, Woodside solutions were 88.74\% as high as the lot-unconstrained problem. Now consider the second row of data, where the only difference to the first row is that the objective function in each of the 132 rebalances is multiplied by $\Theta = 10$. In theory, these problems have the same optimal portfolios. Yet, when we observe the statistics, we see that with $\Theta = 10$, that particular rebalance had an expected return 69.94\% as high as the corresponding lot-unconstrained return, with solutions on average 92.05\% worse. This issue happened because CPLEX faces compounding numerical errors when solving the problem, producing vastly different solutions for different values of $\Theta$\footnote{The instances are publicly available in the form of \texttt{.lp} files so that other researchers can verify these remarks.}. We remind the reader that the initial investment was assumed to be \$1 million; this problem is likely more concerning for larger investments.

Note that as we increase $\Theta$, numerical issues become less notable. However, scaling the objective function is an informal ``workaround'' that works for the specific problem of maximising return. There is no indication that such a simplistic procedure would work with more complex portfolio optimisation models.

With round lots, numerical errors still happen, but less prominently, with an apparent equilibrium for $\Theta \geq 100$. For some instances, though, no solution was produced within 900s (2 when $\Theta = 1\text{,}000$). The average computational time required to solve the problems is higher than with odd lots, with 5 instances not optimally solved when $\Theta = 1\text{,}000$. Most rebalances are solved quickly (as shown by the $90^{\text{th}}$-percentile), but the solver faces computational difficulties somewhat often.

The choice of maximising portfolio return illustrates that a standard application of the Woodside model fails often  even when faced with arguably the simplest portfolio selection problem. For more complex portfolio objectives, it might not even be trivial to conclude whether a decline in the optimal solution value is due to the lots requirement or due to numerical issues.

\subsection{Final remarks}

In summary, absolute formulations are either non-convex, avoid explicitly discounting the portfolio value or assume arbitrary conflicting bounds for linearisation. Relative formulations with lots are less common, with fewer features than absolute formulations. Computational results reported in the respective references suggest that both absolute and relative problems are often intractable. Finally, mixed formulations are either non-convex, ignore costs or, in the case of \citeauthor{woodside2013}, are computationally difficult and prone to numerical errors. Table \ref{table2} summarises these observations. Papers with {\bf conflicting objectives} balance the original portfolio optimisation problem with minimising deviation.

\begin{table}[!ht]
\centering
{\footnotesize
\begin{tabular}{|l|c|c|c|c|c|c|}
\hline
Reference & Formulation
& \begin{tabular}[c]{@{}c@{}}Non\\linear\end{tabular} 
& \begin{tabular}[c]{@{}c@{}}Disc. costs\\omitted\end{tabular}
& \begin{tabular}[c]{@{}c@{}}Arbitrary\\bounds\end{tabular}
& \begin{tabular}[c]{@{}c@{}}Conflicting\\objectives\end{tabular}
& \begin{tabular}[c]{@{}c@{}}Computational\\difficulties\end{tabular} \\
\hline
\cite{speranza1996}     &  Absolute   &   &   & X &   & X \\
\cite{mansini1999}      &  Absolute   &   & X & X &   & X \\
\cite{kellerer2000}     &  Absolute   &   & X & X &   & X \\
\cite{jobst2001}        &  Relative   &   & X &   & X & X \\
\cite{mitra2003}        &  Relative   &   & X &   & X & X \\
\cite{mansini2005}      &  Absolute   &   & X & X &   & X \\
\cite{li2006}           &  Absolute   & X &   & X &   & X \\
\cite{angelelli2008}    &  Absolute   &   & X & X &   & X \\
\cite{lin2008}          &  Mixed      & X & X &   &   & X \\
\cite{bonami2009}       &  Mixed      &   & X &   &   & X \\
\cite{bartholomew2009}  &  Mixed      & X & X &   &   & X \\
\cite{angelelli2012}    &  Absolute   &   & X & X &   & X \\
\cite{baumann2013}      &  Relative   &   & X & X &   & X \\
\cite{woodside2013}     &  Mixed      &   &   & X &   & X \\
\hline          
\end{tabular}
\caption{Summary of works in literature with regards to the issues raised in this section.}
\label{table2}
}
\end{table}

In our judgment, \cite{woodside2013} is the closest in producing realistic trades generically (without a single portfolio utility in mind), provided that the arbitrary bounds do not conflict with the primary objective. However, as shown in the empirical example, it often fails due to numerical errors or computational difficulties. In our proposed two-stage approach, we solve two formulations sequentially: first relative, then absolute. The absolute formulation uses integer variables, but all its constraints are linear. Due to fixed first stage weights, costs can be easily discounted and no arbitrary bounds need to be imposed. 

In terms of computational effort, the second stage focuses only on assets for which trades might occur, instead of the whole asset universe. This considerably reduces problem sizes: in Section \ref{sec:experiments} we show that even with integer and binary variables (due to borrowing costs), reasonably-sized instances can still be solved quickly. Also, since no mixed formulation is employed, numerical errors are much less likely to happen.

One issue with two stages is that it is heuristic in nature: the second stage looks for a lot-feasible portfolio that most closely resembles the first stage optimal solution. The larger the investment value, however, the easier it is to find a close match to the original solution. Moreover, linear single stage approaches do not necessarily ensure optimality: many require arbitrary bounds for linearisation; depending on what is the optimal criteria this may conflict with the primary goal and produce suboptimal solutions. But most importantly, in Section \ref{sec:experiments} we show that the two-stage approach can be efficiently solved, unlike other existing approaches which are often computationally intractable or prone to numerical inaccuracies.

The practical feasibility of the second stage enables us to consider critical aspects of real-life trading overlooked in the portfolio optimisation literature: borrowing costs and futures contracts. Before introducing our framework, we present some basic concepts of futures contracts, together with an empirical example illustrating their potential usefulness.

\section{Key concepts of futures contracts}
\label{sec:futures}

There are several operational differences between equities and futures contracts, some of which need to be addressed in order to optimise portfolios that include both classes. We refer to equities as assets that are traded in stock markets and which incur tangible ownership when purchased, such as stocks and exchange-traded funds. A common implicit assumption in the portfolio optimisation literature is that all assets behave like equities, which may overlook important details. Rather than explaining the concept of a futures contract, tn this section we focus on how to account for these differences in terms of our modelling framework. For a comprehensive overview of the mechanics of futures, we refer the reader to \cite{hull2014}.

When trading futures, the three main operational differences are that:
\begin{itemize}
 \item[(i)] futures have expiration dates,
 \item[(ii)] purchasing futures does not incur ownership of the underlying asset, but a deposit in a margin account is required as guarantee for fulfilling contractual obligations, irrespective of the party promising to {\bf buy} (long) or {\bf sell} (short) the asset, 
 \item[(iii)] positions held are leveraged since required margin accounts are usually a fraction of the underlying contract value.
\end{itemize}

\noindent These differences affect not only the two stages of our approach, but also how data must be prepared.

Since futures expire, historical time series are limited to the duration of the contract (from issuance to expiration). Data vendors often provide a rolled or continuous series, one which automatically switches to the next-contract month when the current most-liquid contract expires. The rolled series provides long-term continuity often required by financial decision models that rely on historical data. Also, since the true investment in futures is the amount deposited in the margin account, the returns of the continuous time series must be multiplied by the leverage level to account for the actual observed returns.

Also, commonly, just before expiration brokers ``cancel out'' most contracts by purchasing equivalent long positions for holders of short positions, and likewise by purchasing equivalent short positions for holders of long positions. At this point, if the investor wishes to maintain an exposure to that contract, it is necessary to purchase the next-month contract. This is called a rollover: the investor is forced to liquidate the entire current position and open a new one, which incurs more transaction costs than simply buying/selling the difference between current and desired holdings. This requirement is modelled in our second stage.

In order to convert proportions to holdings, we also account for the leverage level - which defines the amount to be deposited in the margin account. A further issue is that in futures both long and short positions require a cash deposit, as opposed to equities, where shorting does not (theoretically) require any investment upfront, rather creating a debt towards the lender. With regards to the budget constraint, while the portfolio weight associated with a short position in equities is negative, the weight associated with a short position in futures is positive. In practice, identical short and long positions in a futures contract represent the same positive proportion of the total portfolio. On the other hand, their contribution to portfolio returns is leveraged and negative/positive for short/long positions respectively.

As a general illustration, consider a contract worth $\$10\text{,}000$ in gold. Party \texttt{A} promises to buy $\$10\text{,}000$ from Party \texttt{B} at a later date. Let the required margin deposit be 20\% of the contract value for both parties. Both deposit $\$2\text{,}000$ in their respective margin accounts. Suppose that, in the next day, due to gold price variations a new contract is now worth $\$11\text{,}000$. Even though the contract is not expiring today, the brokers transfer the difference from the margin account of \texttt{B} to the margin account of \texttt{A}. This way, the margin account of \texttt{A} now has $\$3\text{,}000$ and the margin account of \texttt{B} has now fallen short to $\$1\text{,}000$.

Notice that if the margin account value falls below the minimum required value, the broker issues a margin call. In this case \texttt{B} must then add another $\$1\text{,}200$ in funds to its account to bring it to $\$2\text{,}200$, which is 20\% of the current contract value. Likewise, \texttt{A} may optionally withdraw part of its margin account since it is above the required level, making it available for investment in other assets. In practice, pre-expiry cash settlements occur often, hence, just prior to the second stage, we assume that the margin account in any currently held positions are maintained at the exact required level, with the excess being put into/taken from cash reserves.

Note that \texttt{A} had a $\$1\text{,}000$ profit on a $\$2\text{,}000$ investment, or a 50\% return - 5 times the 10\% increase in contract value. Accordingly, \texttt{B} had a -50\% return. If \texttt{B} had decided to short $\$10\text{,}000$ in gold directly, they would have borrowed this amount, sold then at that price and owe $\$10\text{,}000$ to the lender. On the next day, the debt would have grown to $\$11\text{,}000$. The returns are the same (albeit deleveraged), but the proportion held is negative instead of positive.

\subsection{Empirical example}
\label{sec:futuresExample}

Futures are of interest to investors due to many reasons, including liquidity, ease in purchasing short positions, possibility of leverage and generally lower transaction costs. Another major advantage is that futures are available for a wide range of financial instruments, tangible or not, including commodities, interest rates, indices, currencies and others. Including futures in a portfolio can help diversify risk beyond traditional stocks and bonds. In this section we illustrate this through an empirical example.

Let us expand the dataset described in Section \ref{sec:woodsideIssue} with three futures contracts:

\begin{itemize}
  \item Micro E-mini S\&P500 futures\footnote{\url{https://www.cmegroup.com/markets/equities/sp/e-mini-sandp500.html}, last visited $17^{\text{th}}$ September 2024.}, with a lot size of 5,
  \item Gold futures\footnote{\url{https://www.cmegroup.com/markets/metals/precious/gold.html}, last visited $17^{\text{th}}$ September 2024.}, with a lot size of 100,
  \item Mini VIX futures\footnote{\url{https://www.cboe.com/tradable\_products/vix/mini\_vix/}, last visited $17^{\text{th}}$ September 2024.}, with a lot size of 100.
\end{itemize}

\noindent Gold and VIX (Volatility index) are generally negatively correlated with the market.

We simulate a portfolio strategy with the same setup as in Section \ref{sec:woodsideIssue}, but this time we choose optimal portfolio weights with the standard min-variance approach \citep{markowitz1952}. The expected returns and covariance matrices are naively estimated from historical data. We run two different strategies: one with only S\&P500 stocks and another with an asset universe expanded with the three futures above (we assume no leverage). For the sake of illustration here we ignore lots and costs.

\begin{figure}[!ht]
\centering
\includegraphics[width=0.8\textwidth]{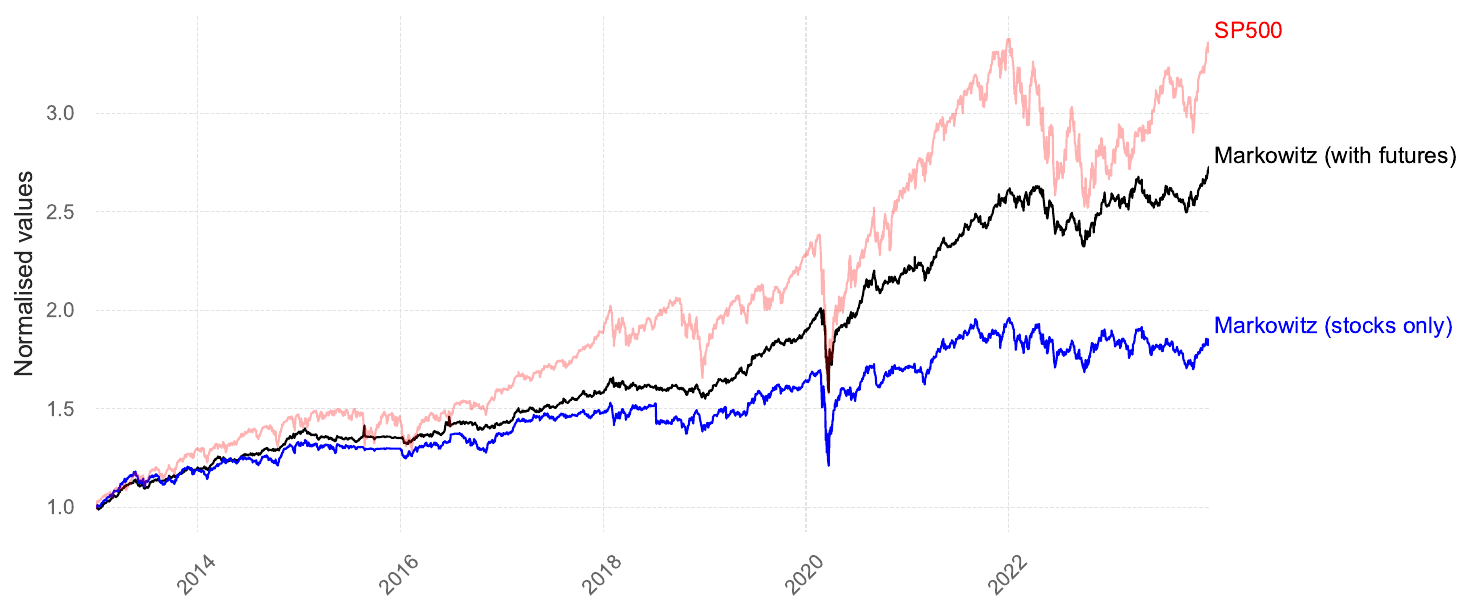}
  \caption{Out-of-sample cumulative performance, minimum variance with and without futures contracts.}
  \label{fig001}
\end{figure}

Figure \ref{fig001} displays the cumulative out-of-sample performance of both strategies, as well as that of the S\&P500. We also show selected out-of-sample statistics in Table \ref{table3}. {\bf FV} stands for the Final Value, {\bf CAGR} for the Compound Annual Growth Rate, {\bf Vol} represents the annualised volatility and {\bf MDD} the maximum drawdown. {\bf Sharpe} and {\bf Sortino} are the annualised Sharpe and Sortino ratios respectively\footnote{The calculation of all statistics is detailed in the appendix.}.

\begin{table}[!ht]
\centering
{\scriptsize
\renewcommand{\tabcolsep}{1mm} 
\renewcommand{\arraystretch}{1.4} 
\begin{tabular}{|l|rrrrrr|}
\hline
\multicolumn{1}{|c|}{Strategies} & \multicolumn{1}{c}{FV} & \multicolumn{1}{c}{CAGR} & \multicolumn{1}{c}{Vol} & \multicolumn{1}{c}{MDD} & \multicolumn{1}{c}{Sharpe} & \multicolumn{1}{c|}{Sortino}\\
\hline
Markowitz (with futures) &     2.72 &     9.54 &     8.54 &    21.23 &     0.84 &     1.23\\
Markowitz (stocks only)  &     1.86 &     5.80 &    11.63 &    28.60 &     0.35 &     0.48\\
SP500                    &     3.36 &    11.65 &    17.24 &    33.92 &     0.59 &     0.82\\
\hline
\end{tabular}
}
\caption{Comparative out-of-sample statistics}
\label{table3}
\end{table}

A naive implementation of the Markowitz min-variance model is widely regarded as failing to produce good out-of-sample portfolios \citep{demiguel2007}. Yet, by simply adding three futures contracts, we observe considerable improvement in the out-of-sample performance, both in terms of return and risk, with Sharpe and Sortino ratios larger than the S\&P500. This illustrates that increasing diversification with selected futures contracts can be beneficial even for a standard strategy. In Section \ref{sec:marketNeutral}, we also show an application of leveraged futures contracts in producing market neutral portfolios with potential for positive return.

\section{First stage: choosing optimal portfolio weights}
\label{sec:firstStage}

The first stage consists simply in solving the portfolio selection model of choice in order to obtain optimal weights. In this section we present a standardised relative formulation that allows us to not only account for futures contracts, but also to incorporate several other constraints. We show here a small selection of constraints, but an extended version is provided in the appendix accompanying this paper.

Let $N$ be the set of $|N|$ assets in which we may invest. We split $N$ into two sets, namely $N_1$ as the set of equities and $N_2$ as the set of futures contracts, with $N_1, N_2 \subseteq N$ and $N_1 \cup N_2 = N$. We informally refer to assets in $N_1$ as ``equities'' and assets in $N_2$ as ``futures''. 

Let $P$ be the current valuation of the portfolio, and let there also be two optional risk-free assets $\ell, b \in N_1$ in which we may invest. Risk-free asset $\ell$ represents the interest rate obtained when we hold a long position in cash (lending money to the bank). Asset $b$ represents the interest rate applied when we hold a short position in cash (borrowing money from the bank). Let:

\vspace{0.3cm}
\begin{tabular}{ll}
$w_i^+ \geq 0, w_i^- \geq 0$ & be the proportion of the portfolio held in long/short positions in asset $i \in N$. \\
$z_i^+, z_i^-$ & \begin{math}\left\{ \begin{array}{ll} 1 & \textrm{if any of asset $i$ is held in long/short positions}\\ 0 & \textrm{otherwise} \end{array} \right. \end{math}
\end{tabular}
\vspace{0.3cm}

\noindent The presence of binary variables $z^+_i$ and $z^-_i$ depends on which exogenous constraints are applicable. Note also that $w_{\ell}^- = 0$ and $w_{b}^+ = 0$.

In portfolio selection literature, the proportion invested in asset $i$ is more often represented with a single variable $w_i$, where $w_i > 0$ means a long position and $w_i < 0$ means a short position. A single variable, however, prevents us from modelling certain classes of exogenous constraints, such as shorting limits in equities \citep{jacobs2005, kumar2010} or short positions in futures.

Let $\rho(\vec{w})$, where $\vec{w} = (w^+_1, w^-_1, \dots, w^+_n, w^-_n)$, be the objective function to be minimised (we assume a minimisation problem without loss of generality). A generic portfolio selection model is given by:
{\small
\begin{optprog}
\optaction[]{min} & \objective{\rho(\vec{w})} \label{eq1} \\
subject to & \sum_{i \in N_1} (w^+_i - w^-_i)  +  \sum_{i \in N_2} (w^+_i + w^-_i)& = & 1 \label{eq2}\\
           & w^+_i & \leq & Uz^+_i & \forall i \in N  \label{eq3} \\
           & w^-_i & \leq & Uz^-_i & \forall i \in N  \label{eq4} \\
           & z^+_i + z^-_i & \leq & 1 & \forall i \in N  \label{eq5} \\
           & z^+_i, z^-_i & \in & \mathbb{B} \label{eq6} \\
           & w^+_i, w^-_i & \geq & 0 & \forall i \in N \label{eq7}
\end{optprog}
}

\noindent Constraint \eqref{eq2} is the budget constraint. For equities, the contribution of $w^-_i, i \in N_1$ towards the budget is negative, while for futures the contribution of $w^-_i, i \in N_2$ is positive: this is due to the fact that shorting in futures requires a (positive) cash deposit in a margin account, unlike shorting in equities. Constraints \eqref{eq3}-\eqref{eq4} link variables $\vec{w}$ and $\vec{z}$. We use the $U$ symbol loosely throughout this text to indicate any large enough constant. Constraints \eqref{eq5} ensure that an asset, if picked, cannot be held in both long and short positions at the same time. Constraints \eqref{eq6}-\eqref{eq7} define variable bounds. Depending on the choice of $\rho(\vec{w})$, the specific formulation may require additional variables and constraints (i.e. minimising CVaR).

Let $\tau^-$, $\tau^+$ be the minimum and maximum proportions of $P$ allowed in a collection of short positions in equities. In order to enforce these limits, we add the following constraint:
{\small
\begin{align}
\tau^- \;\leq\; \sum_{i \in N_1} w^-_i \;& \leq\;  \tau^+ \label{eq8}
\end{align}
}

\noindent With this definition, we replace Constraints \eqref{eq3}-\eqref{eq4} by:
{\small
\begin{align}
w^+_i &  \leq \;(1 + \tau^+) z^+_i & \forall i \in N  \label{eq9} \\
w^-_i & \leq \;(1 + \tau^+) z^-_i & \forall i \in N_2  \label{eq10} \\
w^-_i & \leq \;\tau^+ z^-_i & \forall i \in N_1  \label{eq11}
\end{align}
}

\noindent If $\tau^+ > 0$, then it is possible to invest up to $1 + \tau^+$ in long positions. 


Let $\upsilon^-$, $\upsilon^+$ be the minimum and maximum proportions of $1 + \tau$ allowed in a collection of short positions in futures. To enforce these limits, we add the following constraint:
\begin{equation}
 \left(1 + \sum_{i \in N_1} w^-_i \right) \upsilon^- \;\leq\; \sum_{i \in N_2} w^-_i \;\leq\;  \left(1 + \sum_{i \in N_1} w^-_i \right) \upsilon^+ \label{eq12}
\end{equation}

\noindent We defined these limits as relative to the total long exposure, thus we have that $0 \leq \upsilon^- \leq \upsilon^+ \leq 1$. They could also be optionally defined as absolute values instead.

Let $\mu_i$ be the expected return of asset $i \in N$. In scenario-based formulations, $\mu_i = L_i \sum_{s \in S} p_s r_{is}$, where $L_i$ is the leverage level of asset $i$ ($L_i = 1$ for $i \in N_1$) and $p_s$ is the probability of scenario $s \in S$. If a given futures $i$ has expected return 1\% and leverage level equal to $5 \times$, then effectively $\mu_i = 5\%$.
The following constraint enforces the minimum accepted expected portfolio return given a target return $\bar{\mu}$:
{\small
\begin{align}
 \sum_{i \in N} \mu_i (w_i^+ - w_i^-) & \geq \bar{\mu}
\label{eq17}
\end{align}
}

\noindent Note that short positions in futures contribute negatively to the portfolio expected return.

\section{Second stage: from optimal proportions to real trades}
\label{sec:secondStage}

The solution of the first stage is a vector $\vec{w^*} = (w^*_1, \dots, w^*_n)$ containing the desired portfolio proportions, obtained by calculating $w^*_i = w^{+*}_i - w^{-*}_i$ for all $i \in N$. In this section we introduce linear and mixed-integer formulations to find feasible holdings $\vec{x}$ that match $\vec{w^*}$ as well as possible.

At this stage, the two risk-free assets $\ell$ and $b$ are unnecessary since we know whether $w^*_{\ell} > 0$, $w^*_{b} < 0$ or $w^*_{\ell} = w^*_{b} = 0$. Let $c$ be the index of the cash asset (to which lots and transaction costs do not apply) and let $w^*_c = w^*_{\ell} + w^*_{b}$ be the desired proportion in cash.

We present two versions of the second stage formulation: without lots, and with lots. The former is modelled as a linear program while the latter requires integer variables. Later on we expand both formulations to include borrowing costs - costs associated with short selling shares from assets in $N_1$.

For clarity we redefine some notation introduced in Section \ref{sec:issues}. Let $X_i$ and $V_i$ be the current number of shares/units and the current price of asset $i$. 
Let $M_i = \frac{X_i V_i}{L_i}$ be the financial value invested in asset $i$. In the case of futures, we assume that the margin accounts hold the exact value required to secure the contracts in place, i.e. any surplus to the margin required levels has been added to $X_c$ and any deficit has been taken from $X_c$. We have $P = C + \sum_{i \in N} M_i$ as the current valuation of the portfolio, where in this context $C$ represents available capital to be invested in the portfolio. Note that $C$ might be negative if a cash withdrawal is about to take place. Let $f_i$ be the fractional cost of buying/selling a unit of asset $i \in N$. Parameters $V_c, L_c$ and $f_c$ are undefined. 

Let $\N = c \cup \left\{ i \in N\setminus \{\ell, b\} \mid X_i \neq 0 \text{ or } w^*_i \neq 0 \right\}$ be the set of assets for which trading might occur. Often $|\N| \ll |N|$, which helps in reducing the computational burden of solving the second stage. Let $\N_1 \subseteq \N$ be the set of equities and $\N_2 \subseteq \N$ be the set of futures for which trading might occur, where $\N_1 \cup \N_2 = \N$ and $\N_1 \cap \N_2 = \emptyset$. Let $\N^+ = \left\{ i \in \N \mid w^*_i > 0 \right\}$, $\N^- = \left\{ i \in \N \mid w^*_i < 0 \right\}$ and $\N^0 = \left\{ i \in \N \mid w^*_i = 0 \right\}$. 
Finally, let $\N^R \subseteq \N_2$ be the set of assets which require a rollover, where the current position must be entirely liquidated before purchasing a new one.

\subsection{Second stage formulation, without lots}

Without lots, it is possible to find $\vec{x}$ such that the discounted portfolio perfectly matches $\vec{w}^*$. Hence, in this formulation the objective is to minimise the transaction costs spent to purchase $\vec{x}$. Let us define the following decision variables:

\vspace{0.3cm}
\begin{tabular}{ll}
$p$      & the portfolio value after transaction costs are discounted, \\
$x_{i}$  & number of shares of asset $i \in \N$ to be held, \\
$m_{i}$  & financial value to be invested in asset $i \in \N$, \\
$c$      & amount to be invested in cash, \\
$G_{i}$  & fractional transaction costs associated with buying or selling asset $i \in \N$.
\end{tabular}
\vspace{0.3cm}

\noindent We assume for simplicity that $f_i > 0 \;\; \forall i \in \N$. If $\exists i \mid f_i = 0$, then it is not necessary to define the corresponding $G_i$.

The following linear program finds $\vec{x}$ while spending as little as possible in transaction costs:

{\small
\begin{optprog}
\optaction[]{min} &  \objective{ \sum_{i \in \N} G_i \label{eqTC1}}\\
subject to & c    & = & w^*_c p &   \label{eqTC2} \\
& m_i  & = & w^*_i p,      &  i \in \N_1  \cup \left( \N_2 \cap \left(\N^+ \cup \N^0\right) \right) \label{eqTC3} \\
& m_i  & = & -w^*_i p,      & i \in \N_2 \cap \N^- \label{eqTC4} \\
& m_i  & = & \frac{V_i x_i}{L_i},                        &  i \in \N_1  \cup \left( \N_2 \cap \left(\N^+ \cup \N^0\right) \right) \label{eqTC5}\\
& m_i  & = & - \frac{V_i x_i}{L_i},                      &  i \in \N_2 \cap \N^- \label{eqTC6}\\
& G_i  & \geq & (x_i - X_i) f_i V_i,                   &  i \in \N \label{eqTC7}
\end{optprog}
\begin{optprog}
& G_i  & \geq & (X_i - x_i) f_i V_i,                   &  i \in \N \label{eqTC8} \\
& G_i  & \geq & (- x_i - X_i) f_i V_i,                 &  i \in \N^R  \label{eqTC9}\\
& G_i  & \geq & (X_i + x_i) f_i V_i,                   &  i \in \N^R \label{eqTC10} \\
& p & = & P - \sum_{j \in \N} G_j \label{eqTC11}
\end{optprog}
}

Constraints \eqref{eqTC2}-\eqref{eqTC4} ensure that the financial value held in each asset corresponds to the exact desired proportions. Constraints \eqref{eqTC5}-\eqref{eqTC6} calculate the necessary margin accounts (or financial valuation) invested in each asset. The $m_i$ variables are not strictly necessary but are kept to ease the mathematics presented. Constraints \eqref{eqTC7}-\eqref{eqTC10} calculate fractional transaction costs, including when rollover applies. Coupled with the ``downward pressure" from Objective \eqref{eqTC1}, they ensure that the $G_i$ variables are equal to the costs spent. Finally, Constraint \eqref{eqTC11} defines the discounted portfolio value. This constraint is also not strictly necessary, but simplifies the presentation of the other formulations below.

Due to space reasons we do not consider piecewise convex or concave costs, nor fixed costs. These could be easily introduced, however, with the addition of extra binary variables. We leave this for future work.

\subsection{Second stage formulation, with lots}

When integral lots are required, it is generally impossible to find a valid portfolio that perfectly matches the desired proportions. We propose a goal programming approach that, within the set of lot-feasible portfolios, finds the one that minimises the deviation to the desired weights.

We do, however, still want to minimise the transaction costs spent, so there are two potentially conflicting objectives. On one hand, minimising the deviations to $\vec{w^*}$ may require spending more in transaction costs, on the other hand saving on transaction costs imply keeping the portfolio close to its current form, potentially deviating more than necessary from $\vec{w^*}$. We propose a weighted sum of both objectives. We adopt the policy that minimising the deviation takes precedence over minimising transaction costs, and discuss appropriate weights for the objective function.

Let $\N^L$ and $\N^{\centernot{L}}$ be the set of assets for which lots apply/do not apply, with $\N^L \cup \N^{\centernot{L}} = \N$. Let us redefine $X_i$ as the current number of shares/units of asset $i \in \N^{\centernot{L}}$ or the current number of lots of asset $i \in \N^L$. Likewise, we redefine $V_i$ as the current price of one unit of asset $i \in \N^{\centernot{L}}$ or one lot of asset $i \in \N^L$.

With regards to the cash asset $c$, a negative deviation to $w^*_c$ may imply a portfolio with higher risk, which may not be acceptable. Therefore we also propose an optional control parameter $W_c$ as the lower accepted limit on the actual proportion invested in cash. If $w^*_c \geq 0$, then having $W_c = 0$ ensures that the actual portfolio does not borrow any cash. If $W_c = w^*_c$ (irrespective of the sign of $w^*_c$), then we ensure that the proportion in cash will be at least $w^*_c$, i.e. which ensures that if $w^*_c < 0$ no extra cash is borrowed. We make use of the following decision variables:

\vspace{0.3cm}
\begin{tabular}{ll}
$x_{i} \in \mathbb{Z}$ & redefined as number of lots of $l_i$ shares of asset $i \in \N^L$ \\
$x_{i} \in \mathbb{R}$ & as number of units of assets $i \in \N^{\centernot{L}}$ \\
$d^+_c, d^-_c \geq 0$ & as the upward and downward deviation from $w^*_c$ in the final portfolio. \\
$d^+_i, d^-_i \geq 0$ & as the upward and downward deviation from the implied financial value of the number \\
                      & of shares that perfectly represents the proportion $w^*_i, i \in \N$ of the final portfolio.
\end{tabular}
\vspace{0.3cm}

\noindent Deviations are expressed as financial values so as to combine them with costs into a single objective function.

Let $\N_1^- = \{i \in \N_1 \mid X_i < 0\}$ be the set of assets in $\N_1$ for which we currently hold short positions. The second stage formulation with lots is given by:
{\small
\begin{optprog}
\optaction[]{min} &  \objective{d^+_c + d^-_c + \sum_{i \in \N} (d^+_i + d^-_i) + \sum_{i \in \N} W_i G_i\label{eqLOT1}}\\

subject to & \eqref{eqTC5}-\eqref{eqTC11} & & & \nonumber \\
  & \sum_{i \in \N} m_i + c & =    & p,                &   \label{eqLOT2}\\
&  c & \geq & W_c p,  & \label{eqLOT3}\\
&  c + d^-_c - d^+_c & =    & w^*_c p,  & \label{eqLOT4}\\
& m_i + d^-_i - d^+_i & = & w^*_i p &  i \in \N_1  \cup \left( \N_2 \cap \left(\N^+ \cup \N^0\right) \right) \label{eqLOT5} \\
& m_i + d^-_i - d^+_i & = & -w^*_i p &  i \in \N_2 \cap \N^- \label{eqLOT6} \\
& \sum_{\mathclap{i \in \N^-_1}} m_i & \geq  & \sum_{\mathclap{i \in \N^-_1}} w^*_i p  & \label{eqLOT7} \\
& x_i & \geq & 0 &  i \in \N^+ \label{eqLOT8} \\
& x_i & \leq & 0 &  i \in \N^- \label{eqLOT9} \\
& x_i & = & 0 &  i \in \N \setminus \{\N^+ \cup \N^-\} \label{eqLOT10}
\end{optprog}
}

\noindent The two conflicting goals in Objective \eqref{eqLOT1} are combined with weights $W_i$. Constraint \eqref{eqLOT2} are balance constraints that calculate the discounted portfolio value. Constraint \eqref{eqLOT3} enforces the minimum allowed proportion in cash, while Constraints \eqref{eqLOT4}-\eqref{eqLOT6} calculate the deviations from $\vec{w^*}$. Optional Constraint \eqref{eqLOT7} prevents overexposure in short positions. Constraints \eqref{eqLOT8}-\eqref{eqLOT10} ensure that assets whose desired position is long (short) are not held in short (long) positions, and that assets whose desired exposure is zero are not to be held in any residual amounts.

\subsubsection{Objective function}

We mentioned previously the assumption that minimising the deviation takes precedence over minimising the transaction costs required to purchase the new portfolio. Since in Objective \eqref{eqLOT1} all terms are expressed as financial values, we may calculate $W_i$ such that the cost improvement in reducing deviation is not offset by the increase in transaction costs spent. Note that different fractional transaction costs may apply to different asset classes or even assets, such as when we artificially increase $f_i$ as a penalty for illiquid assets.

We standardise the definition of the objective function weights by letting $W_i = \theta \frac{1}{f_i L_i}, 0 < \theta \leq 1$, for $i \in \N$. If $\theta = 1$, then reducing the deviation by \$1 incurs exactly \$1 in transaction costs penalties. For leveraged assets, the cost is paid with regards to the full contract, but the deviation is reduced according to the margin account required deposit, hence the $L_i$ term in the denominator. If $\theta = 0$, then a trivial optimal solution is any such that $p = \sum_{i \in N} G_i$. In that case, $c = 0$, $x_i = 0, i \in N$ and all deviations are trivially zero.

If we employ $\theta \gg 1$, transaction costs are heavily penalised and the optimal solution tends to trade as little as possible. As $\theta \to 0$, optimal solutions artificially inflate transaction costs (meaning $G_i > |x_i - X_i|$) in order to minimise deviations, excessively discounting $p$. In Section \ref{sec:objFunctionEvaluation}, we discuss appropriate values of $\theta$.

\subsection{Comparison to \cite{woodside2013}}
\label{sec:comparison}

In this section we compare our two-stage approach to the Woodside model, following the same experimental setup discussed in Section \ref{sec:woodsideIssue}: maximising expected return subject to a 5\% upper bound per asset. Here in all test instances there are more than 20 assets with positive expected return, hence in the abscence of numerical errors any optimal Woodside solution is the true lot-feasible optimal portfolio. Following the results in Table \ref{table1},
to reduce the likelihood of numerical problems we set $\Theta = 1 \times 10^4$.

In our framework, we first find the lot-unrestricted optimal portfolio with our first stage and then find the lot-feasible portfolio that minimises the total deviation to the optimal weights. The first stage formulation is given in the appendix and the second stage formulation is obtained by minimising \eqref{eqLOT1} subject to \eqref{eqTC5}-\eqref{eqTC11} and \eqref{eqLOT2}-\eqref{eqLOT10}. For a fair comparison, we also include the following additional constraints in our second stage:
{\small
\begin{align}
m_i & \leq 0.05 p &  i \in \N_1  & \label{eqLOT11}
\end{align}
}

\noindent Without them, the second stage may produce a portfolio that invests above the upper limit in assets with the highest expected returns.

The higher the financial value involved, the easier it is to approximate the lot-unrestricted optimal weights. In order to verify this we vary the initial investment from $\$100\text{,}000$ to $\$5$ million. We display comparative results between Woodside and the second stage in-sample solutions in Table \ref{table4}.

\begin{table}[!ht]
\centering
{\scriptsize
\renewcommand{\tabcolsep}{1mm} 
\renewcommand{\arraystretch}{1.4} 
\begin{tabular}{|l|l|rrrr|rrr|rrrrrrr|}
\hline
\multicolumn{1}{|c|}{\multirow[c]{2}{*}{Lots}} & \multicolumn{1}{c|}{\multirow[c]{2}{*}{\begin{tabular}{c}Initial\\investment\end{tabular}}} & \multicolumn{4}{c|}{Woodside} & \multicolumn{3}{c|}{Second stage} & \multicolumn{7}{c|}{Difference}\\
\cline{3-6}
\cline{7-9}
\cline{10-16}
 &  & \multicolumn{1}{c}{Success} & \multicolumn{1}{c}{Min} & \multicolumn{1}{c}{Avg} & \multicolumn{1}{c|}{Max} & \multicolumn{1}{c}{Min} & \multicolumn{1}{c}{Avg} & \multicolumn{1}{c|}{Max} & \multicolumn{1}{c}{Min} & \multicolumn{1}{c}{$P_{25}$} & \multicolumn{1}{c}{Avg} & \multicolumn{1}{c}{Median} & \multicolumn{1}{c}{$P_{75}$} & \multicolumn{1}{c}{Max} & \multicolumn{1}{c|}{$< 0$}\\
\hline
\multirow[c]{5}{*}{Odd} & 100k     &    132 &  99.37 &  99.87 &  99.96 &  98.78 &  99.52 &  99.82 &   0.10 &   0.23 &   0.35 &   0.30 &   0.43 &   1.03 &      0\\
 & 200k     &    132 &  99.80 &  99.93 &  99.98 &  99.08 &  99.76 &  99.92 &   0.03 &   0.10 &   0.16 &   0.14 &   0.20 &   0.86 &      0\\
 & 500k     &    132 &  99.91 &  99.96 & 100.00 &  99.79 &  99.91 &  99.97 &  -0.01 &   0.02 &   0.05 &   0.04 &   0.07 &   0.19 &      7\\
 & 1M       &    132 &  99.93 &  99.97 & 100.00 &  99.83 &  99.95 &  99.98 &  -0.04 &   0.00 &   0.02 &   0.01 &   0.03 &   0.15 &     33\\
 & 5M       &    132 &  99.82 &  99.97 & 100.00 &  99.97 &  99.99 & 100.00 &  -0.17 &  -0.03 &  -0.02 &  -0.02 &  -0.01 &   0.02 &    116\\
\cline{1-16}
\multirow[c]{5}{*}{Round} & 100k     &    127 &  67.84 &  83.79 &  95.82 &  29.73 &  51.40 &  79.07 &  12.64 &  25.81 &  32.43 &  33.04 &  38.94 &  51.01 &      0\\
 & 200k     &    131 &  84.79 &  93.30 &  98.47 &  45.84 &  74.56 &  89.56 &   6.12 &  13.14 &  18.77 &  17.96 &  23.69 &  40.17 &      0\\
 & 500k     &    130 &  92.18 &  97.67 &  99.25 &  80.29 &  90.36 &  96.73 &   2.52 &   4.94 &   7.28 &   6.66 &   9.45 &  15.88 &      0\\
 & 1M       &    131 &  92.52 &  98.82 &  99.68 &  87.78 &  95.03 &  98.41 &   1.23 &   2.48 &   3.79 &   3.45 &   4.74 &   9.52 &      0\\
 & 5M       &    132 &  98.41 &  99.76 &  99.92 &  97.00 &  99.06 &  99.64 &   0.23 &   0.45 &   0.70 &   0.63 &   0.85 &   2.82 &      0\\
\hline
\end{tabular}
}
\caption{Comparison of expected return decline (against the first stage lot-unconstrained optimal solutions) between the Woodside model and the second stage for various values of the initial investment. $P_{\alpha}$ denotes the $\alpha^{\text{th}}$ percentile.}
\label{table4}
\end{table}

Under label Woodside, the {\bf Success} column shows the number of rebalances where the solver found a feasible solution within 900s. The other columns compare Woodside with the lot-unrestricted first-stage solutions. The same data is shown for the second stage under the label {\bf Second stage}, noting that the computational time required to solve both stages in sequence was never higher than 0.5 seconds.
We then take, for each rebalance, the difference between the declines observed in Woodside and the second stage, and show corresponding statistics under label {\bf Difference}. Only successful Woodside rebalances are included in this calculation. The last column, labelled $\mathbf{< 0}$, shows the number of times the Woodside solution was actually {\bf worse} than the second stage solution. Since all expected returns are positive this can be either caused by numerical inaccuracies or suboptimal Woodside solutions. We verified, however, that numerical errors account for all of them since Woodside found the optimal in all rebalances where this happened.

Table \ref{table1} had shown that, with odd lots, the Woodside model was more prone to numerical inaccuracies. Even with $\Theta = 1 \times 10^4$, we can also see this happen in Table \ref{table4}: with $\$5$ million as initial investment, in 116 rebalances the Woodside solution was worse than the second stage. Round lots are more restrictive and (according to Table \ref{table1}) make Woodside less prone to numerical errors. When the initial investment is small ($\$100\text{,}000$), the benefits of using a single-stage approach are observed more clearly: the Woodside expected returns were on average $32.43$\% closer to the first stage than the second stage solutions. This difference reduces gradually until, with $\$5$ million, second stage solutions are on average only $0.7\%$ worse.

From these results, we can conclude that there is clear a case where employing a single stage approach is advantageous. These experiments, however, are built to ``help'' Woodside as much as possible: under simplified conditions (the choice of portfolio model), where we could mitigate the issue of numerical errors. Moreover, the Woodside model is often slow or, in some cases, does not even find any feasible solutions. Under more realistic conditions we believe Woodside is even more likely to fail. Additionally, the higher the initial investment, the more likely the second stage is to find solutions as good as those from Woodside {\bf when} the latter is able to find those solutions. In this sense, the second stage can be employed as a tool for judging the quality of the single stage approach.

\section{Borrowing costs}
\label{sec:borrowingCosts}

Short positions in equities require the borrower to pay a ``rent fee'' to the lender. This fee is paid whenever a shorting contract is closed or expires. Borrowing costs are generally defined as a preset yearly rate to be paid proportionally to the time the contract was held, with a reference price as basis for the calculation (usually the closing price of the day before the contract was opened). The actual borrowing rate depends on the supply of stocks for shorting, and varies both among assets and over time.

Consider the following example. We initiate a shorting contract at day 0 where we borrow 200 shares of some asset at \$20.00 each. Suppose the rent rate at the time was 5\% a year, and that we liquidate the contract after 10 business days. Under the assumptions that the reference price is the contract initial price and that the calculations are done with compounded returns, the actual costs to be paid are:
{\small
\begin{align*}
200 \times \$20 \times \left( (1 + 0.05)^{\frac{10}{252}} - 1\right) & \approx \$7.75 
\end{align*}
}

\noindent where 252 approximates the number of business days in a year. The calculation details depend on the broker and the stock market.

We may also partially close a contract. Consider that we are closing 100 shares in 10 days, and the other 100 shares in 15 days. We pay:
{\small
\begin{align*}
\text{On day 10: }  100 \times \$20 \times \left( (1 + 0.05)^{\frac{10}{252}} - 1\right) & \approx \$3.88 \\
\text{On day 15: }  100 \times \$20 \times \left( (1 + 0.05)^{\frac{15}{252}} - 1\right) & \approx \$5.82
\end{align*}
}

\noindent which make up a total of approximately \$9.69.

The agreed rate is specific to each individual contract. For instance, we may currently hold a total short position of 300 shares, but we may have initiated 100 shares 10 days ago (Contract 1), 100 shares 5 days ago (Contract 2) and another 100 shares 2 days ago (Contract 3). The rates agreed for each contract may differ: let the rate for Contract 1 be 3\%, the rate for Contract 2 be 5\% and the rate for Contract 3 be 4\%. For each of these contracts, costs are calculated individually.

When partially closing a short position, we need to decide which of the contracts should be closed. Suppose in the example above that we want to liquidate 150 shares and keep another 150. We may:

\begin{itemize}
    \item Close them chronologically, effectively liquidating Contract 1 in its entirety (paying its borrowing costs) and 50 shares of Contract 2 (again, paying its partial costs).
    \item Close the most expensive first, so liquidating Contract 2 in its entirety, then partially liquidating 50 shares of Contract 3 while keeping Contract 1 unchanged.
\end{itemize}

When rebalancing towards a given portfolio $\vec{w}^*$, we might have to close some short positions. On top of fractional transaction costs, borrowing costs might have to be deducted, impacting the calculation of the discounted portfolio value $p$. Since borrowing costs are entirely based on a predefined rate, the contract reference price and the number of days since the contract started, it is possible to calculate prior to the optimisation how much we would pay if an individual shorting contract was closed. In the case of partial closures, we may also sort the individual contracts based on which we would like to close first.

For assets in $\N_1^-$, let us break down $X_i$ into a sum of multiple open shorting contracts. Let $J_i$ be the number of open short contracts of asset $i$, and let $X^j_i$, $j = 1, \ldots, J_i$ (such that $\sum_{j = 1}^{J_i} X^j_i = X_i$), define the amount held originating from each contract, sorted from the most desirable to close first to the least desirable. Let $h^j_i$ be the monetary value to be paid in borrowing costs if we fully closed contract $X^j_i$ today, with every $h_i^j$ being calculated prior to the optimisation. For instance, if $X^j_i = -100$, and the corresponding contract has a reference price of \$20 and a rate of 4\%, 7 business days ago, then $h^j_i =  100 \times \$20 \times \left( (1 + 0.04)^{\frac{7}{252}} - 1\right) \approx \$2.18$.

\subsection{Formulation with borrowing costs, no lots}

Let $H^j_i : i \in \N_1^-, j = 1, \ldots, J_i$ be additional decision variables representing the costs paid in fully or partially closing contract $j$ from asset $i$. Let $\delta^j_i$, $j = 0, \ldots, J_i + 1$, be auxiliary binary variables such that:

\vspace{0.3cm}
\begin{tabular}{ll}
$\delta^0_i = 1$ & if $x_i \leq X_i$, that is, we either keep or increase our short position in $i$, \\
$\delta^j_i = 1,\; j = 1, \ldots, J_i$ & if $\sum_{k = 1}^{j-1} X^j_i \leq x_i - X_i \leq \sum_{k = 1}^{j} X^j_i$, that is, if we close all contracts up to $j-1$, \\ 
                                       & and partially or entirely close contract $j$. If $j = 1$, $\sum_{k = 1}^{j-1} X^j_i = 0$.\\
$\delta^{J_i+1}_i = 1$, & if $x_i \geq 0$, that is, if we close the entire short position in $i$.
\end{tabular}
\vspace{0.3cm}

The second stage formulation with borrowing costs and no lots is defined as:
{\small
\begin{optprog}
\optaction[]{min} &  \objective{ \sum_{i \in \N} G_i + \sum_{\mathclap{i \in \N^-_1}} \sum_{j = 1}^{J_i} H^j_i\label{eqTCR1}}\\
subject to & \eqref{eqTC2}-\eqref{eqTC10} & & & \nonumber \\
& p & = & P - \sum_{i \in \N} G_i - \sum_{\mathclap{i \in \N^-_1}} \sum_{j = 1}^{J_i}H^j_i \label{eqTCR2}  \\
& \sum_{j = 0}^{\mathclap{J_i + 1}} \delta^j_i & = & 1 &  i \in \N^-_1 \label{eqTCR3}\\
& x_i - X_i & \leq & U(1 - \delta^0_i) &   i \in \N^-_1 \label{eqTCR4} \\ 
&  x_i - X_i &  \geq & -U(1 - \delta^1_i) &  i \in \N^-_1 \label{eqTCR5} \\
&  x_i - X_i & \leq & -\sum_{k = 1}^{j}X^j_i + U(1 - \delta^j_i) &   i \in \N^-_1, j = 1, \ldots, J_i \label{eqTCR6}  \\
&  x_i - X_i & \geq & -\sum_{k = 1}^{j-1}X^j_i - U(1 - \delta^j_i) &   i \in \N^-_1, j = 2, \ldots, J_i+1 \label{eqTCR7} 
\end{optprog}
\begin{optprog}
& H_i^j & \geq & 0 &   i \in \N^-_1, j = 1, \ldots, J_i \label{eqTCR8} \\
& H_i^j & \geq & h^j_i \frac{x_i - (X_i - \sum_{k = 1}^{j-1}X^k_i )}{-X^j_i} - U \left(\sum_{k = j+1}^{J_i+1} \delta^k_i \right) & i \in \N^-_1, j = 1, \ldots, J_i \label{eqTCR9} \\
& H_i^j & \geq & h^j_i - h_i^j \left(\sum_{k = 0}^{j} \delta^k_i \right) & i \in \N^-_1, j = 1, \ldots, J_i \label{eqTCR10}
\end{optprog}
}

Constraint \eqref{eqTCR2} calculates the discounted portfolio $p$ with deducted borrowing costs. Constraints \eqref{eqTCR3}-\eqref{eqTCR7} ensure that the appropriate binary variable $\delta_i^j, j = 0, \dots, J_i+1$, is correctly assigned. Constant $U$ can be calculated as the maximum possible difference $| x_i - X_i |$ (depending on $w^*_i$). Constraints \eqref{eqTCR8}-\eqref{eqTCR10}, together with Objective \eqref{eqTCR1}, calculate the borrowing costs to be paid per contract.

\subsection{Formulation with borrowing costs and lots}

Let $f^j_i = - \frac{h^j_i}{X^j_i V_i}$ be the proportion of the value of the short contract $j$ from asset $i$ to be paid in borrowing costs in case we fully close it. Let us redefine $W_i$, $i \in \N^-_1$ in such a way that reducing the deviation by \$1 does not incur more than a \$1 penalty in variable plus borrowing costs altogether. A natural approach to enforce this is to combine $f^j_i$ with $f_i$ in a single proportion - however, since $f^j_i$ is contract dependent, we must define $G_i^j$ and $W^j_i$ instead of $G_i$ and $W_i$, for $j = 0, \ldots, J_i + 1, i \in \N^-_1$. More specifically, for $0 < \theta \leq 1$:
\begin{itemize}
\item $W_i^0 = \theta \frac{1}{f_i}$, with $G^0_i > 0$ representing fractional costs associated with an increase in the short position currently held, so no contracts being closed.
\item $W_i^{J_i + 1} = \theta \frac{1}{f_i}$, with $G^{J_i + 1}_i > 0$ representing fractional costs associated with opening a long position in $i$, if required.
\item $W_i^j = \theta \frac{1}{\left( f^j_i + f_i\right)}, j = 1, \ldots, J_i$, with $G^j_i + H^j_i$ representing the sum of fractional and borrowing costs paid when closing, or partially closing, contract $j$.
\end{itemize}

\noindent Note that we have ignored the term $L_i$ since for $i \in \N^-_1$, $L_i = 1$.

For $i \in \N^-_1$, let
\vspace{-0.5cm}
{\small
\begin{align*}
\Upsilon(i) = W_i^0 G_i^0 + W_i^{J_i + 1} G_i^{J_i+1} + \sum_{j = 1}^{J_i} W_i^j \left(H^j_i + G_i^j\right).
\end{align*}
}

\noindent Let also $g^j_i = - f_i V_i X_i^j$ represent the fractional costs paid if the whole contract $j$ is closed. The second stage formulation with lots and borrowing costs is given by:
{\small
\begin{optprog}
\optaction[]{min} &  \objective{d^+_c + d^-_c + \sum_{i \in \N} (d^+_i + d^-_i) + \sum_{\mathclap{\substack{i \in \N \setminus \N^-_1}}} W_i G_i + \sum_{i \in \N^-_1} \Upsilon(i)\label{eqTCRL1}}\\

subject to & \eqref{eqTC5}-&\eqref{eqTC6}&, \eqref{eqTC9}-\eqref{eqTC10}, \eqref{eqLOT2}-\eqref{eqLOT10}, \eqref{eqTCR2}-\eqref{eqTCR10}  \nonumber \\
& G_i                    & \geq & (x_i - X_i) f_i V_i,                   &  i \in \N \setminus \N^-_1  \label{eqTCRL2}\\
& G_i                    & \geq & (X_i - x_i) f_i V_i,                   &  i \in \N \setminus \N^-_1\label{eqTCRL3}\\
& G_i & = & \sum_{\mathclap{j = 0}}^{J_i+1} G_i^j  & i \in \N^-_1 \label{eqTCRL4} \\
& G_i^j & \geq & 0 & i \in \N^-_1, j = 0, \ldots, J_i+1 \label{eqTCRL5} \\
& G_i^j & \geq & g^j_i \frac{x_i - (X_i - \sum_{k = 1}^{j-1}X^k_i )}{-X^j_i} - U \left(\sum_{k = j+1}^{J_i+1} \delta^k_i \right) & i \in \N^-_1, j = 1, \ldots, J_i \label{eqTCRL6} 
\end{optprog}
\begin{optprog}
& G_i^j & \geq & g^j_i - g_i^j \left(\sum_{k = 0}^{j} \delta^k_i \right) & i \in \N^-_1, j = 1, \ldots, J_i \label{eqTCRL7} \\
& G_i^0 & \geq & f_i V_i (X_i - x_i) & i \in \N^-_1 \label{eqTCRL8} \\
& G_i^{J_i+1} & \geq & f_i V_i x_i  & i \in \N^-_1 \label{eqTCRL9}
\end{optprog}
}

Constraints \eqref{eqTCRL2}-\eqref{eqTCRL9} enforce the newly defined $G_i^j$ variables. More specifically, Constraints \eqref{eqTCRL6}-\eqref{eqTCRL7} enforce that the fractional costs for any single contract are only nonzero if the corresponding contract is partially or fully closed.

\section{Computational experiments}
\label{sec:experiments}

On top of producing (hopefully) investable portfolios, and since the framework is agnostic to the portfolio optimisation model of choice, the second stage may also be employed {\bf against} candidate financial strategies by highlighting any difficulties an investor would face before investing (and potentially losing!) real money. By incorporating multiple costs and accumulating portfolio discounts in subsequent rebalances, we get closer to the ``true'' profit that would have been achieved by the strategy. Even if indirectly, in our view this possibility also helps in constructing a ``bridge'' between model and investment.

In order to do so, the second stage must quickly and reliably produce holdings that match the desired portfolio as accurately as possible given the realistic market features applicable. Hence, in this section, our focus is in evaluating the second stage itself, with regards to its applicability, computational time and solution quality. We discuss limitations and related insights of the second stage in Section \ref{sec:managerial}. With regards to the first stage, the main difference to the literature is the inclusion of futures contracts, which give more liberty to fund managers in hedging and leveraging portfolios. In Section \ref{sec:marketNeutral} we present a case study illustrating these features.

In our custom-built simulation tool, when a portfolio rebalance takes place, the optimisation framework invokes the first stage and then the second stage. Afterwards, it ``executes'' the recommended transactions, discounting the necessary costs. Interest in cash is accrued every day, according to the 10-year US treasury bond (symbol TNX) returns. This series is used as a synthetic asset representing the return on both lent and borrowed cash. We use here the expanded dataset described in Section \ref{sec:futuresExample}, with stocks and futures.

In the S\&P500 dataset we deal with around 500 assets at any point in time, which is a reasonable amount of assets when operating an investment fund. Surely it is possible to expand this asset universe to include larger indices (Russel 2000, Russel 3000) or to include small-caps. Compiling this data however is not trivial without access to specialised data providers (i.e. finding changes in indices compositions spanning many years). We leave the evaluation of the two-stage framework with larger asset universes for future work, but we believe the current dataset represents a realistic scenario.

Apart from being part of the index at the time (with the exception of futures and cash), an asset is only considered eligible in a rebalance if there are at most 5\% prices missing from the most recent historical series of 201 prices (200 returns). Any missing prices in the series of any asset that passes this filter is then completed with forward fill, then backwards fill. Whenever a simulation has costs involved, we assume 5 basis points as variable transaction costs and 3\% a year as borrowing costs in equities. The out-of-sample and in-sample periods as well as the rebalancing frequency are the same as stated in Section \ref{sec:woodsideIssue}. Unless explicitly noted we simulate an initial investment of $\$500\text{,}000$.

Due to futures contracts rollover, the second stage is executed more often than the first stage. Whenever a rollover happens on a non-rebalancing day, we invoke the second stage in order to obtain a portfolio discounted by the rollover costs and which is as close as possible to the currently held portfolio. The time-series for every futures contract is the front-month contract with automatic rollover. Rollover dates for each contract are taken from their corresponding websites. The leverage level of each contract depends on the experiment.

All the data here described is publicly available at:
\begin{center}
\url{https://github.com/cristianoarbex/portfolioSimulationData}.
\end{center}

\subsection{Case study}
\label{sec:caseStudy}

We begin by presenting a case study of our framework. The goal here is to evaluate whether the second stage works ``against the strategy'' and whether it is ``practical'' with regards to computational time and solution depreciation (when compared to that obtained in first stage). In later sections we evaluate the last two specific points in more detail.

As investment strategy we employ the \cite{valle2017b} implementation of the second-order stochastic dominance (SSD) model for enhanced indexation proposed by \cite{roman2013}. The goal here is to take a seemingly promising strategy and evaluate it when forced to abide by market rules, in this case by adding lots, costs and rollover. We report here both long only and 140/40 long/short strategies (where $\tau = 0.4$). We impose lower and upper bounds of $5\%$ and $20\%$ applicable to any joint investment in the set $Q$ of derivatives, and we set $L_i = 1$ for all $i \in Q$ (no leverage). When lots are applicable, in the second stage we set $W_c = w^*_c$ and $\theta = 0.05$.

We run three versions of both strategies: (i) without costs and lots (so no second stage needed), (2) with costs and odd lots in stocks and (3) with costs and round lots in stocks. For futures contracts, we use the lots described in Section \ref{sec:futuresExample} for both (2) and (3). We map the strategies to their required second stage formulations in Table \ref{table5}.

\begin{table}[!ht]
\centering
\renewcommand{\tabcolsep}{1mm} 
\renewcommand{\arraystretch}{1} 
{\footnotesize
\begin{tabular}{|l|l|l|}
\hline
Strategy & $2^{\text{nd}}$ stage formulation & Costs applicable \\
\hline
Long only               &  Not necessary  &  None  \\
Long only (odd lots)    &  $\min \eqref{eqLOT1} \text{ s.t. } \eqref{eqTC5}\text{-}\eqref{eqTC11}, \eqref{eqLOT2}\text{-}\eqref{eqLOT10}$  &  Fractional \\
Long only (round lots)  &  $\min \eqref{eqLOT1} \text{ s.t. } \eqref{eqTC5}\text{-}\eqref{eqTC11}, \eqref{eqLOT2}\text{-}\eqref{eqLOT10}$  &  Fractional  \\
Long/short              &  Not necessary  &  None  \\
Long/short (odd lots)   &  $\min \eqref{eqTCRL1} \text{ s.t. } \eqref{eqTC5}\text{-}\eqref{eqTC6}, \eqref{eqTC9}\text{-}\eqref{eqTC10}, \eqref{eqLOT2}\text{-}\eqref{eqLOT10}, \eqref{eqTCR2}\text{-}\eqref{eqTCR10}, \eqref{eqTCRL2}\text{-}\eqref{eqTCRL9}$  &  Fractional and borrowing  \\
Long/short (round lots) &  $\min \eqref{eqTCRL1} \text{ s.t. } \eqref{eqTC5}\text{-}\eqref{eqTC6}, \eqref{eqTC9}\text{-}\eqref{eqTC10}, \eqref{eqLOT2}\text{-}\eqref{eqLOT10}, \eqref{eqTCR2}\text{-}\eqref{eqTCR10}, \eqref{eqTCRL2}\text{-}\eqref{eqTCRL9}$  &  Fractional and borrowing \\
\hline          
\end{tabular}
\caption{Strategies in the case study and their respective required second stage formulations.}
\label{table5}
}
\end{table}

\begin{figure}[!ht]
\centering
\includegraphics[width=0.85\textwidth]{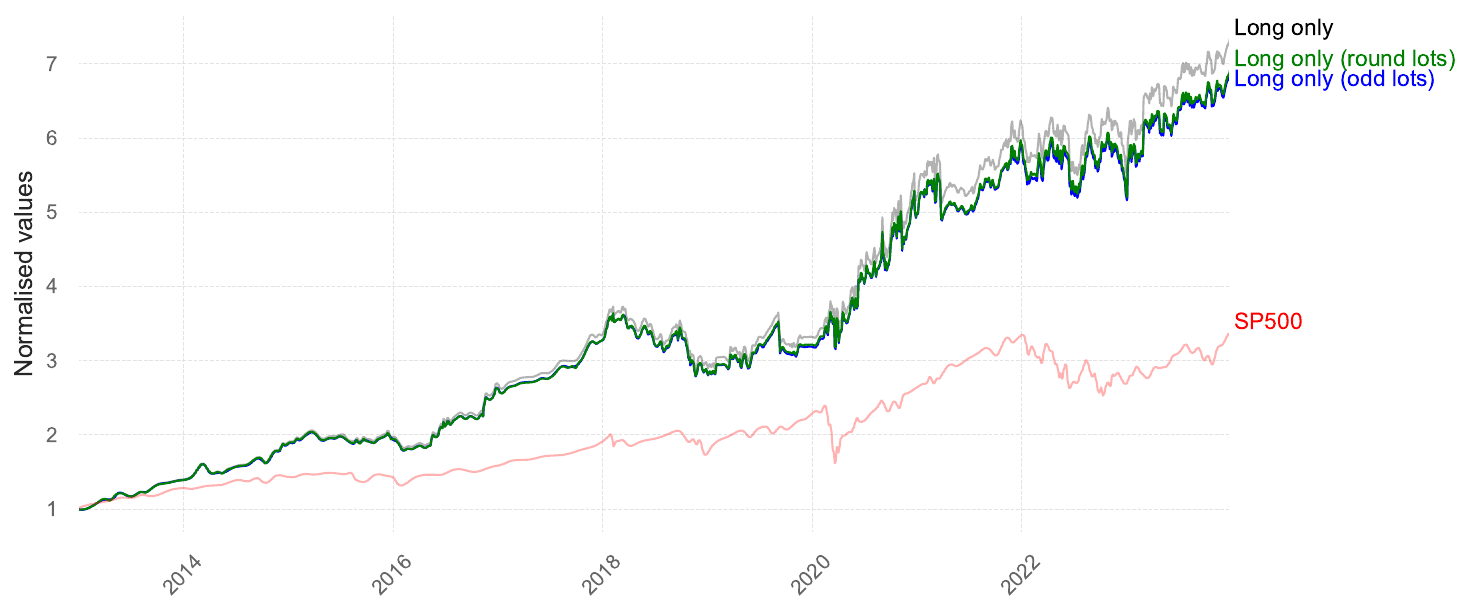}
  \caption{Out-of-sample cumulative performance, long only experiments with and without lots/costs.}
  \label{fig002}
\end{figure}

\begin{figure}[!ht]
\centering
\includegraphics[width=0.85\textwidth]{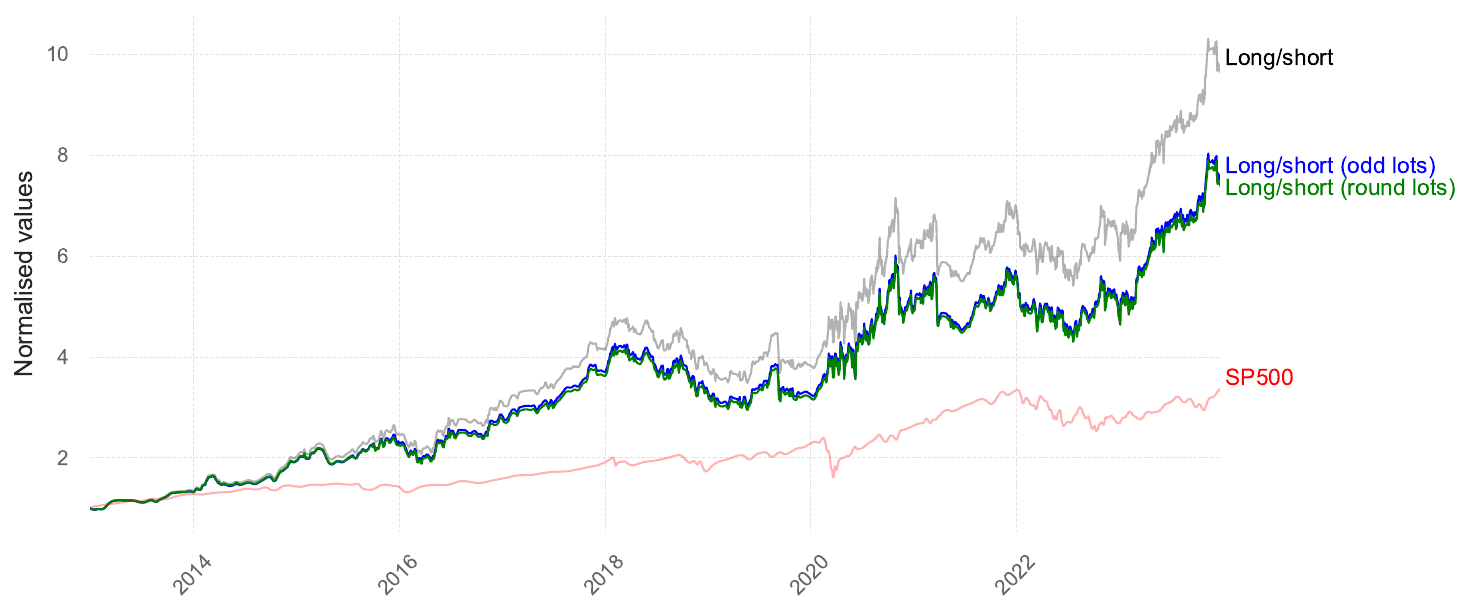}
  \caption{Out-of-sample cumulative performance, long/short experiments with and without lots/costs.}
  \label{fig003}
\end{figure}

We display the cumulative out-of-sample returns of all strategies and that of the S\&P500 in Figures \ref{fig002} and \ref{fig003}. The lines in the charts were smoothed with univariate splines for improved visualisation. Additionally, Table \ref{table6} shows selected out-of-sample statistics (described in Section \ref{sec:futuresExample}) for all six experiments and the S\&P500. 

\begin{table}[!ht]
\centering
{\scriptsize
\renewcommand{\tabcolsep}{1mm} 
\renewcommand{\arraystretch}{1.4} 
\begin{tabular}{|l|rrrrrr|}
\hline
\multicolumn{1}{|c|}{Strategies} & \multicolumn{1}{c}{FV} & \multicolumn{1}{c}{CAGR} & \multicolumn{1}{c}{Vol} & \multicolumn{1}{c}{MDD} & \multicolumn{1}{c}{Sharpe} & \multicolumn{1}{c|}{Sortino}\\
\hline
Long only               &     7.33 &    19.85 &    17.39 &    24.24 &     1.00 &     1.50\\
Long only (odd lots)    &     6.85 &    19.12 &    17.39 &    24.68 &     0.96 &     1.45\\
Long only (round lots)  &     6.90 &    19.20 &    17.36 &    24.45 &     0.97 &     1.46\\
Long/short              &     9.70 &    22.94 &    22.75 &    27.46 &     0.92 &     1.36\\
Long/short (odd lots)   &     7.55 &    20.17 &    22.75 &    29.66 &     0.82 &     1.21\\
Long/short (round lots) &     7.43 &    20.00 &    22.71 &    29.69 &     0.82 &     1.20\\
SP500                   &     3.36 &    11.65 &    17.24 &    33.92 &     0.59 &     0.82\\
\hline
\end{tabular}
}
\caption{Comparative out-of-sample statistics}
\label{table6}
\end{table}

Without costs and lots, both selected strategies vastly outperform the S\&P500 in terms of performance (FV, CAGR) and risk-adjusted performance (Sharpe, Sortino). In the long only strategy, the only costs involved are fractional costs when buying and selling shares and those associated with the rollover of futures contracts. With regards to performance, the addition of odd lots and costs resulted in a decrease of 0.73\% in CAGR, a $\approx4\%$ decrease in Sharpe and $\approx3\%$ in Sortino ratios. When round lots are employed we are likely to observe worse second stage approximations of the first stage weights. Yet in this particular case the out-of-sample performance was similar to that with odd lots, with a 0.65\% drop in CAGR and a $\approx3\%$ drop in the Sharpe and Sortino ratios.

Borrowing costs are applicable in the long/short strategies, which further impact the strategy performance. With odd lots we observe a decrease of 2.77\% in CAGR, 11\% in the Sharpe and Sortino ratios, while with round lots we observe drops of 2.94\% in CAGR, 11\% in the Sharpe ratio and 12\% in the Sortino ratio. In fact, the long/short performance with round lots only slightly outperforms the corresponding long only strategy, but with much higher risk (Vol, MDD) and hence worse risk-adjusted returns. This performance however is likely no worse than what would have been observed in practice since the asset universe under consideration is sufficiently liquid for the relatively small initial investment. Moreover, the borrowing and fractional costs applied possibly overestimate the true costs that would have been charged by brokers.

With regards to the second stage applicability, an important consideration is the computational time required to solve it, especially when borrowing costs are included. This is critical since in a live trading environment, once a decision is made regarding portfolio weights, the system responsible for trading has to wait until the second stage is solved and weights are converted into trades. Table \ref{table7} shows statistics related to the number of trades, portfolio cardinality (number of assets with non-zero weights) and the computational time required to solve each of the 132 second stage instances (those due to portfolio rebalances). We remind the reader that the second stage is executed after each rebalance as well as rollover (non-rebalancing) days; the calculation of statistics however includes only those due to rebalances. In general the second stage execution due to rollover is faster as fewer trades are required, which could skew the statistics positively.

\begin{table}[!ht]
\centering
{\scriptsize
\renewcommand{\tabcolsep}{1mm} 
\renewcommand{\arraystretch}{1.4} 
\begin{tabular}{|l|rrrr|rrrr|rrrr|}
\hline
\multicolumn{1}{|c|}{\multirow[c]{2}{*}{Strategies}} & \multicolumn{4}{c|}{Number of trades} & \multicolumn{4}{c|}{Portfolio cardinality} & \multicolumn{4}{c|}{Computational time (s)}\\
\cline{2-13}
 & \multicolumn{1}{c}{Min} & \multicolumn{1}{c}{Avg} & \multicolumn{1}{c}{Median} & \multicolumn{1}{c|}{Max} & \multicolumn{1}{c}{Min} & \multicolumn{1}{c}{Avg} & \multicolumn{1}{c}{Median} & \multicolumn{1}{c|}{Max} & \multicolumn{1}{c}{Min} & \multicolumn{1}{c}{Avg} & \multicolumn{1}{c}{Median} & \multicolumn{1}{c|}{Max}\\
\hline
Long only (odd lots)    &      5 &  19.94 & 18.00 &   43 &      4 &  14.61 & 14.00 &   33 &   0.01 &   0.03 &   0.02 &   0.10\\
Long only (round lots)  &      5 &  18.02 & 17.00 &   37 &      4 &  14.61 & 14.00 &   33 &   0.01 &   0.02 &   0.02 &   0.07\\
Long/short (odd lots)   &     14 &  45.44 & 46.50 &   89 &      9 &  32.64 & 32.50 &   69 &   0.02 &   0.25 &   0.17 &   1.50\\
Long/short (round lots) &     14 &  41.81 & 43.00 &   82 &      9 &  32.64 & 32.50 &   69 &   0.02 &   0.19 &   0.14 &   1.12\\
\hline
\end{tabular}
}
\caption{Number of trades, portfolio cardinality and second stage computational time per strategy}
\label{table7}
\end{table}

From the table we observe that the long only strategies required fewer trades and had less diversified portfolios in general when compared to the long/short strategies. Under these conditions, the second stage computational time is negligible. The long/short strategies had more diversified portfolios, more trades and binary variables due to borrowing costs; yet in the worst case the second stage required only 1.5s to be solved to optimality. If this case study represents a relatively realistic setting for an investment fund, then in such case the second stage could be incorporated into a live trading system without major concerns. 

Another important aspect in the second stage is the solution quality in terms of deviation from the desired first stage proportions. Table \ref{table8} shows, for each experiment, selected statistics regarding the total deviation, per rebalance, from the corresponding desired proportions. The total deviation is the sum of the absolute values of all individual asset deviations. As an illustration, if the desired portfolio was 120\% in asset A, -20\% in asset B and 0\% in cash, and the real portfolio invests 117\% in A, 2\% in cash and -19\% in B, then the total deviation is 6\%.

\begin{table}[!ht]
\centering
{\scriptsize
\renewcommand{\tabcolsep}{1mm} 
\renewcommand{\arraystretch}{1.4} 
\begin{tabular}{|l|rrrrrrrr|}
\hline
\multicolumn{1}{|c|}{Strategies} & \multicolumn{1}{c}{Min} & \multicolumn{1}{c}{$P_{10}$} & \multicolumn{1}{c}{$P_{25}$} & \multicolumn{1}{c}{Avg} & \multicolumn{1}{c}{Median} & \multicolumn{1}{c}{$P_{75}$} & \multicolumn{1}{c}{$P_{90}$} & \multicolumn{1}{c|}{Max}\\
\hline
Long only (odd lots)    &   0.01 &   0.03 &   0.05 &   0.09 &   0.07 &   0.10 &   0.16 &   0.54\\
Long only (round lots)  &   0.53 &   1.12 &   1.68 &   3.12 &   2.67 &   3.90 &   5.77 &  10.48\\
Long/short (odd lots)   &   0.02 &   0.04 &   0.07 &   0.40 &   0.09 &   0.16 &   0.43 &   5.70\\
Long/short (round lots) &   1.42 &   2.44 &   3.52 &   5.98 &   5.59 &   7.55 &  10.39 &  17.33\\
\hline
\end{tabular}
}
\caption{Total deviations (in \%), per rebalance, from desired proportions. Selected statistics, where $P_{\alpha}$ denotes the $\alpha^{\text{th}}$ percentile.}
\label{table8}
\end{table}

Non surprisingly, Table \ref{table8} shows that total deviations under the more strict round lot policy are further away from the desired first stage proportions. With odd lots, the average total deviation was only 0.09\% and 0.40\% for long only and long/short strategies. The median was even lower, at 0.07\% and 0.09\%, which suggests the presence of scattered outliers among the 132 rebalances. With round lots, not only the averages are higher (3.12\% and 5.98\%), but also the medians are relatively closer to the mean (2.67\% and 5.59\%). In the worst case, we observed around 17\% total deviation. We emphasise, however, that this is the sum of deviations of around 20-80 trades (as observed in Table \ref{table7}), so still not a high value overall.


\subsubsection{Effect of initial investment}
\label{sec:investmentAndLots}

\begin{figure}[!ht]
\centering
\includegraphics[width=0.8\textwidth]{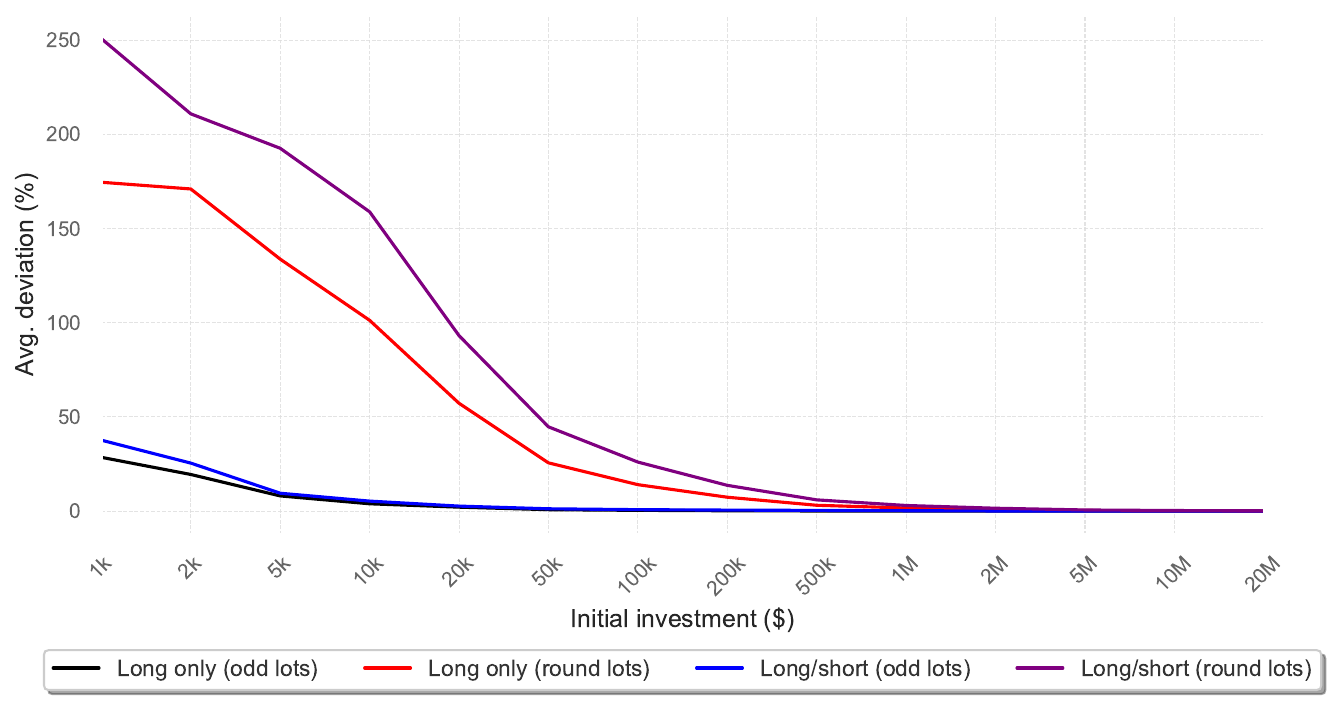}
  \caption{Average total deviation as a function of the initial investment.}
  \label{fig004}
\end{figure}

The goal programming nature of the second stage means that the smaller the portfolio value is, the less likely the second stage model is to find solutions that are close to the desired proportions. In this section, we vary the initial investment of both the long only and long/short strategies of this case study. We show in Figure \ref{fig004} the average deviation per rebalance as a function of the initial investment.

For small investment values, such as $\$1\text{,}000$, we observe very large average deviations since a single lot of most assets exceeds the investment proportion allocated to it. Specifically, we observe average deviations of approximately $250\%$ and $175\%$ average deviation for the long/short and long only strategies with round lots. However, as the initial investment amount increases, average deviations decrease substantially. For an investment of $\$20\text{,}000\text{,}000$, reasonable for small to medium institutional funds, the average total deviation for the long/short strategy with round lots reduces to just $0.17\%$. As the financial amount increases, the behaviour of the second stage becomes less heuristic, approaching the exact lot-unconstrained solutions.

\subsection{Second stage, computational time}

In the previous section, the computational time required to solve the second stage was negligible. In this section, we force more diversification and more trades in order to evaluate its execution time under more demanding scenarios. In order to do that we add an upper exposure bound $u$ per asset to the long only and long/short strategies of Section \ref{sec:caseStudy}. We also increase the initial investment capital from $\$500\text{,}000$ to $\$50$ million to more easily diversify the portfolios, especially when lots are considered. We do not allow investment in cash, so as to prevent cash from taking a large proportion of the portfolios. We ignore futures rollover in order to focus only on the 132 rebalances. We also changed the threshold of accepted percentage of invalid prices from 5\% to 20\% with the goal of increasing the size of the asset universe. For the optimisation of the second stage, we set a time limit of 300s per rebalance. 

We simulate all lot policies: none (no lots), odd and round, with costs applicable to all of them. A summary of the experiments is given in Table \ref{table9}, where we show the formulations employed in each case and the values of $u$ considered. We also show the types of decision variables required in each formulation. We did not employ an upper bound of 0.25\% in the long/short strategies as it results in many infeasible problems. 

Table \ref{table10} shows the results. Statistics about cardinality and the number of trades are a good indicator of instance sizes: the larger these numbers are, the larger the second stage model is. Computational times are displayed next, where column {\bf TL} shows in how many of the 132 instances the solver reached the time limit. For these instances, the solution employed in the simulation is the best solution found as the solver is halted. Finally, the last set of columns shows statistics about optimality gaps among the unsolved instances - cases where the solver found the optimal within 300s {\bf are not} included.

\begin{table}[!ht]
\centering
\renewcommand{\tabcolsep}{1mm} 
\renewcommand{\arraystretch}{1} 
{\footnotesize
\begin{tabular}{|l|l|l|l|l|}
\hline
Strategy & Lots & $2^{\text{nd}}$ stage formulation & Values of $u$ & Variable types\\
\hline
 \multirow[c]{3}{*}{Long only} & None & $\min \eqref{eqTC1} \text{ s.t. } \eqref{eqTC2}\text{-}\eqref{eqTC11}$  & 1\%, 0.5\%, 0.25\%  & Fractional \\
                               & Odd & $\min \eqref{eqLOT1} \text{ s.t. } \eqref{eqTC5}\text{-}\eqref{eqTC11}, \eqref{eqLOT2}\text{-}\eqref{eqLOT10}$  & 1\%, 0.5\%, 0.25\% & Integer\\
                               & Round & $\min \eqref{eqLOT1} \text{ s.t. } \eqref{eqTC5}\text{-}\eqref{eqTC11}, \eqref{eqLOT2}\text{-}\eqref{eqLOT10}$  &   1\%, 0.5\%, 0.25\% & Integer \\
\hline
\multirow[c]{3}{*}{Long/short} &  None & $\min \eqref{eqTCR1}  \text{ s.t. }  \eqref{eqTC2}\text{-}\eqref{eqTC10}, \eqref{eqTCR2}\text{-}\eqref{eqTCR10}$ &  2\%, 1\%, 0.5\% & Binary \\
                               &  Odd & $\min \eqref{eqTCRL1} \text{ s.t. } \eqref{eqTC5}\text{-}\eqref{eqTC6}, \eqref{eqTC9}\text{-}\eqref{eqTC10}, \eqref{eqLOT2}\text{-}\eqref{eqLOT10}, \eqref{eqTCR2}\text{-}\eqref{eqTCR10}, \eqref{eqTCRL2}\text{-}\eqref{eqTCRL9}$  & 5\%, 2\%, 1\%, 0.5\% & Integer and binary\\
                               &  Round & $\min \eqref{eqTCRL1} \text{ s.t. } \eqref{eqTC5}\text{-}\eqref{eqTC6}, \eqref{eqTC9}\text{-}\eqref{eqTC10}, \eqref{eqLOT2}\text{-}\eqref{eqLOT10}, \eqref{eqTCR2}\text{-}\eqref{eqTCR10}, \eqref{eqTCRL2}\text{-}\eqref{eqTCRL9}$  & 5\%, 2\%, 1\%, 0.5\% & Integer and binary\\
\hline          
\end{tabular}
\caption{More diversified strategies for evaluating the computational time required to solve the $2^{\text{nd}}$ stage.}
\label{table9}
}
\end{table}

\begin{table}[!ht]
\centering
{\scriptsize
\renewcommand{\tabcolsep}{1mm} 
\renewcommand{\arraystretch}{1.4} 
\begin{tabular}{|l|l|l|rrr|rrr|rrrr|rrrr|}
\hline
\multicolumn{1}{|c|}{\multirow[c]{2}{*}{Strategy}} & \multicolumn{1}{c|}{\multirow[c]{2}{*}{Lots}} & \multicolumn{1}{c|}{\multirow[c]{2}{*}{$u$}} & \multicolumn{3}{c|}{Cardinality} & \multicolumn{3}{c|}{Num. Trades} & \multicolumn{4}{c|}{Computational time (s)} & \multicolumn{4}{c|}{Optimality gap (\%)}\\
\cline{4-17}
 &  &  & \multicolumn{1}{c}{Min} & \multicolumn{1}{c}{Avg} & \multicolumn{1}{c|}{Max} & \multicolumn{1}{c}{Min} & \multicolumn{1}{c}{Avg} & \multicolumn{1}{c|}{Max} & \multicolumn{1}{c}{Min} & \multicolumn{1}{c}{Avg} & \multicolumn{1}{c}{Max} & \multicolumn{1}{c|}{TL} & \multicolumn{1}{c}{Min} & \multicolumn{1}{c}{Avg} & \multicolumn{1}{c}{$P_{90}$} & \multicolumn{1}{c|}{Max}\\
\hline
\multirow[c]{9}{*}{Long only} & \multirow[c]{3}{*}{None} & 1.0\%    &    100 &    103 &    109 &    103 &    130 &    147 &    0.0 &    0.0 &    0.0 &      0 & --     & --     & --     & --    \\
 &  & 0.5\%    &    200 &    202 &    207 &    202 &    238 &    259 &    0.1 &    0.1 &    0.1 &      0 & --     & --     & --     & --    \\
 &  & 0.25\%   &    400 &    402 &    408 &    403 &    425 &    441 &    0.3 &    0.3 &    0.3 &      0 & --     & --     & --     & --    \\
\cline{2-17}
 & \multirow[c]{3}{*}{Odd} & 1.0\%    &    100 &    103 &    109 &    103 &    130 &    147 &    0.2 &    0.6 &    1.3 &      0 & --     & --     & --     & --    \\
 &  & 0.5\%    &    200 &    202 &    207 &    202 &    237 &    259 &    2.9 &    5.9 &   12.9 &      0 & --     & --     & --     & --    \\
 &  & 0.25\%   &    400 &    402 &    408 &    395 &    422 &    439 &   24.3 &  234.6 & TL     &     83 &   0.01 &   0.05 &   0.13 &   0.29\\
\cline{2-17}
 & \multirow[c]{3}{*}{Round} & 1.0\%    &    100 &    103 &    109 &    102 &    122 &    145 &    0.7 &    1.3 &    2.2 &      0 & --     & --     & --     & --    \\
 &  & 0.5\%    &    200 &    202 &    207 &    176 &    205 &    243 &    9.3 &   16.3 &   29.6 &      0 & --     & --     & --     & --    \\
 &  & 0.25\%   &    400 &    402 &    408 &    247 &    305 &    402 &   56.8 &  202.6 & TL     &     44 &   0.01 &   0.14 &   0.32 &   1.77\\
\cline{1-17}
\multirow[c]{11}{*}{Long/short} & \multirow[c]{3}{*}{None} & 2.0\%    &     90 &     98 &    119 &     96 &    128 &    151 &    0.0 &    0.1 &    0.1 &      0 & --     & --     & --     & --    \\
 &  & 1.0\%    &    180 &    185 &    202 &    183 &    232 &    264 &    0.0 &    0.3 &    0.8 &      0 & --     & --     & --     & --    \\
 &  & 0.5\%    &    360 &    364 &    372 &    362 &    423 &    457 &    0.2 &    2.4 &    7.4 &      0 & --     & --     & --     & --    \\
\cline{2-17}
 & \multirow[c]{4}{*}{Odd} & 5.0\%    &     36 &     47 &     76 &     45 &     65 &     96 &    0.1 &    0.3 &    1.8 &      0 & --     & --     & --     & --    \\
 &  & 2.0\%    &     90 &     98 &    119 &     96 &    128 &    150 &    0.3 &   71.2 & TL     &     22 &   0.01 &   0.01 &   0.02 &   0.02\\
 &  & 1.0\%    &    180 &    185 &    202 &    183 &    231 &    261 &    2.5 &  170.2 & TL     &     68 &   0.01 &   0.04 &   0.07 &   0.08\\
 &  & 0.5\%    &    360 &    364 &    372 &    362 &    422 &    455 &   62.3 &  266.4 & TL     &     96 &   0.01 &   0.08 &   0.17 &   0.30\\
\cline{2-17}
 & \multirow[c]{4}{*}{Round} & 5.0\%    &     36 &     47 &     76 &     44 &     64 &     96 &    0.1 &    0.8 &    8.5 &      0 & --     & --     & --     & --    \\
 &  & 2.0\%    &     90 &     98 &    119 &     96 &    125 &    149 &    1.3 &   38.0 & TL     &      8 &   0.05 &   0.29 &   0.61 &   1.26\\
 &  & 1.0\%    &    180 &    185 &    202 &    183 &    220 &    254 &    5.8 &  180.1 & TL     &     62 &   0.01 &   1.21 &   2.86 &   5.83\\
 &  & 0.5\%    &    360 &    364 &    372 &    327 &    373 &    428 &  300.0 &  300.0 & TL     &    132 &   0.01 &   6.26 &  13.98 &  18.88\\
\hline
\end{tabular}
}
\caption{$2^{\text{nd}}$ stage optimisation statistics for 132 rebalances across different strategies, where $P_{\alpha}$ denotes the $\alpha^{\text{th}}$ percentile.}
\label{table10}
\end{table}

As an illustration, among the long only experiments with round lots and $u = 0.25\%$, the SSD model chose on average non-zero weights in 402 different assets. The second stage suggested on average 305 trades to be executed, with a minimum of 247 and a maximum of 402. The solver could not prove optimality within 300s in 44 rebalances. The $90^{\text{th}}$ percentile optimality gap among these 44 instances was 0.32\%, and in the worst case it was 1.77\%. These particular instances are considerably larger than the case study from Section \ref{sec:caseStudy}.

\begin{figure}[!ht]
\centering
\includegraphics[width=0.8\textwidth]{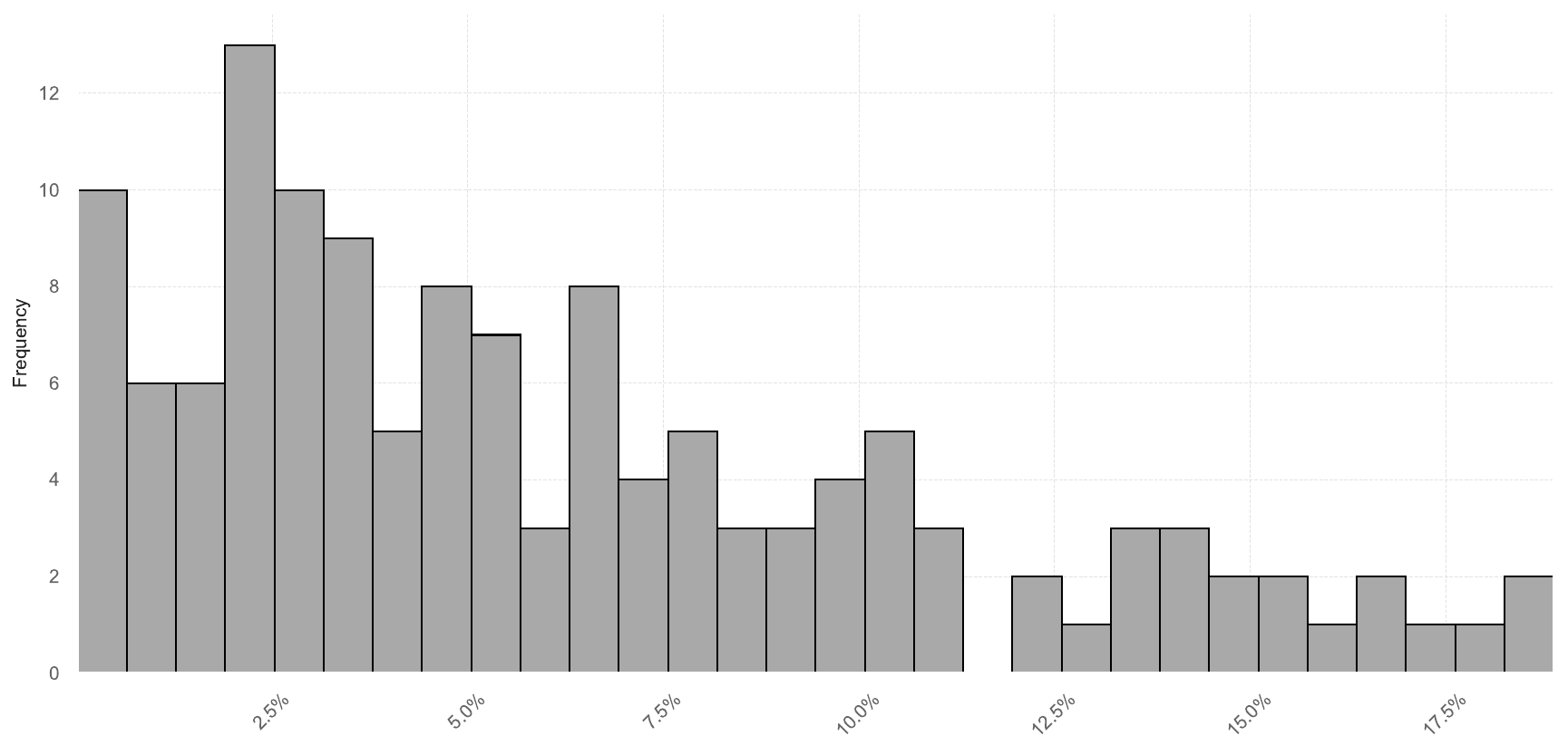}
  \caption{Optimality gaps distribution among 132 unsolved rebalances, long/short experiments with round lots and $u = 0.5\%$.}
  \label{fig005}
\end{figure}

Based on the data presented in the table, we draw the following conclusions:

\begin{itemize}
 \item Without lots (rows labelled {\bf None}), no general integer variables are necessary. In this case the long only models are linear and the computational times are negligible. The long/short experiments on the other hand require binary variables to model borrowing costs. Yet, all 396 rebalances (for three values of $u$) were solved quickly, with an average of 2.4s for the $u = 0.5\%$ and a worst case of 7.4s.

 \item The long only experiments with both lot policies require general integer variables, but no binary variables for borrowing costs. Computational times grow considerably: for $u = 0.25\%$, 127 instances were not solved to optimality. Yet, among the unsolved instances, optimality gaps were very small, with averages of 0.05\% and 0.14\% for odd and round lots. An outlier ended with a 1.77\% gap, but the second largest gap (not reported in the table) was 0.56\%.

 \item The hardest instances are those that include round lots and borrowing costs. The optimality gaps among unsolved long/short instances with odd lots have a worst case of 0.3\%. With round lots and $u = 0.5\%$, however, no instance was solved within 300s. The average gap was 6.26\% and the $90^{\text{th}}$ percentile was 13.98\%, with a worst case of 18.88\%. Figure \ref{fig005} shows the whole distribution of gaps. Despite the gaps being more concentrated below 5\%, at this point the second stage performance starts to deteriorate. As future work we intend to study combinatorial optimisation techniques in order to reduce the computational effort for these hardest cases.

\end{itemize}

To summarise, in this section we enforced upper bounds in order to artificially increase the size of portfolios. The goal was to increase the difficulty of solving the second stage in order to evaluate how it behaves in more extreme conditions. We imposed a time limit of 300s, and the vast majority of instances were either solved much quicker than the limit or were very close to optimal. In practice, for large portfolios the worst case means solving the first stage, waiting 5 minutes and then executing the trades - whether this is realistic or not depends on the context since the current prices are likely to be changed from the prices considered in the second stage. We discuss this further in Section \ref{sec:managerial}.

\subsection{Evaluation of the second stage objective function}
\label{sec:objFunctionEvaluation}

As previously discussed, Objectives \eqref{eqLOT1} and \eqref{eqTCRL1} combine two goals by minimising a weighted sum of deviation and costs. In order to compute these weights, parameter $\theta$ standardises potentially distinct costs among different assets, and is defined as the relative importance of minimising costs as compared to minimising deviation. As we adopt the policy that minimising deviation takes precedence, we employ $\theta < 1$.

In this section, we assess how different values of $\theta$ influence both deviation and costs. Within the formulation, fractional and borrowing costs are held in nonnegative variables $G_i$ and $H_i^j$ respectively, constrained by inequalities. When the inequalities are binding, these variables hold the exact required costs for moving from the current portfolio to the next. However, since minimising costs tends to be conflicting with minimising deviation, there is no assurance that at the optimal solution all inequalities are binding. A very small value of $\theta$ increases the chance of non-binding inequalities (surpluses), while a large value of $\theta$ might increase overall deviation. In all experiments reported in this paper, we allocated eventual surpluses into cash.

\begin{figure}[!ht]
\centering
\includegraphics[width=0.8\textwidth]{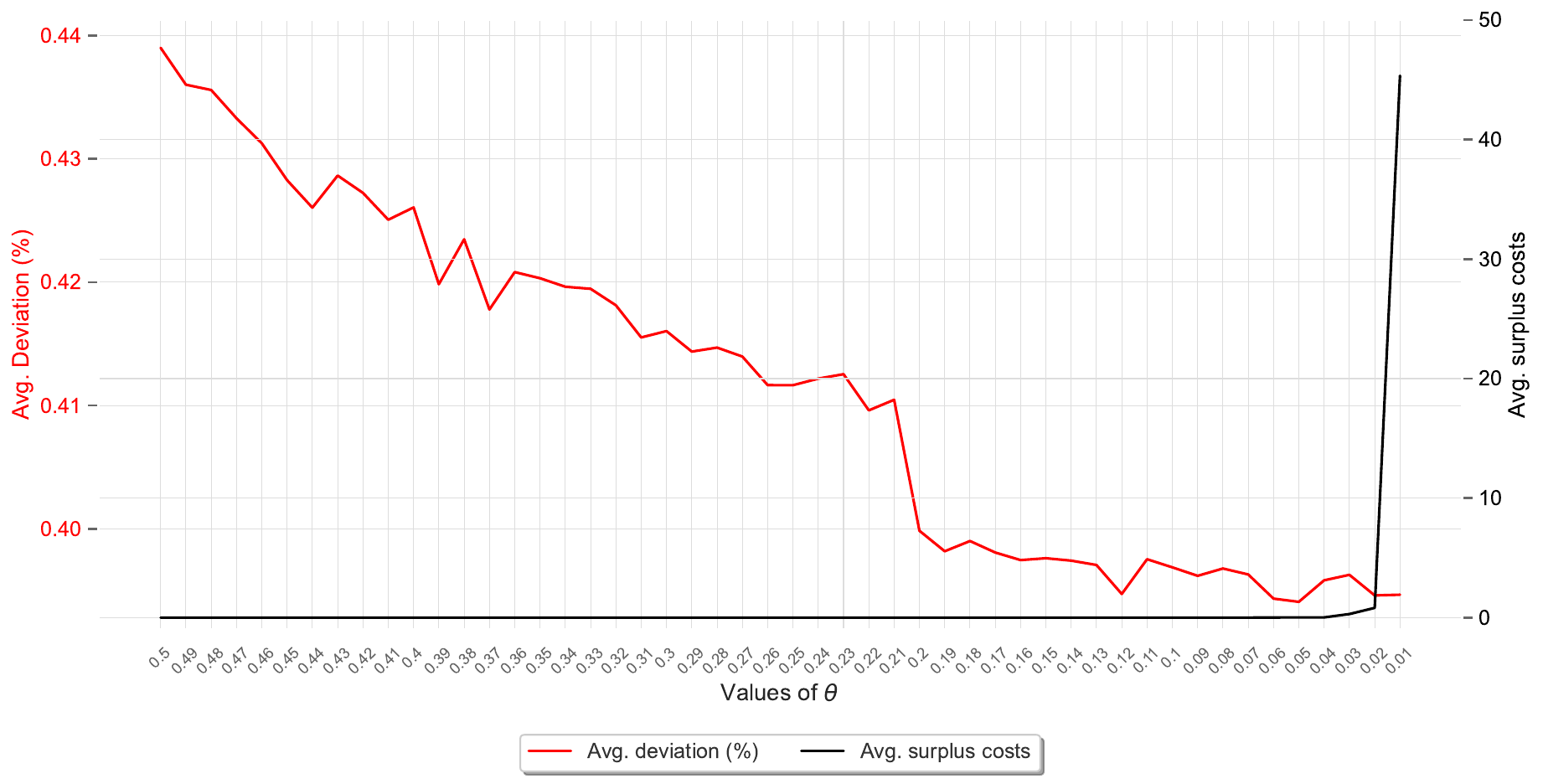}
  \caption{Long/short with odd lots, various values of $\theta$.}
  \label{fig008}
\end{figure}

\begin{figure}[!ht]
\centering
\includegraphics[width=0.8\textwidth]{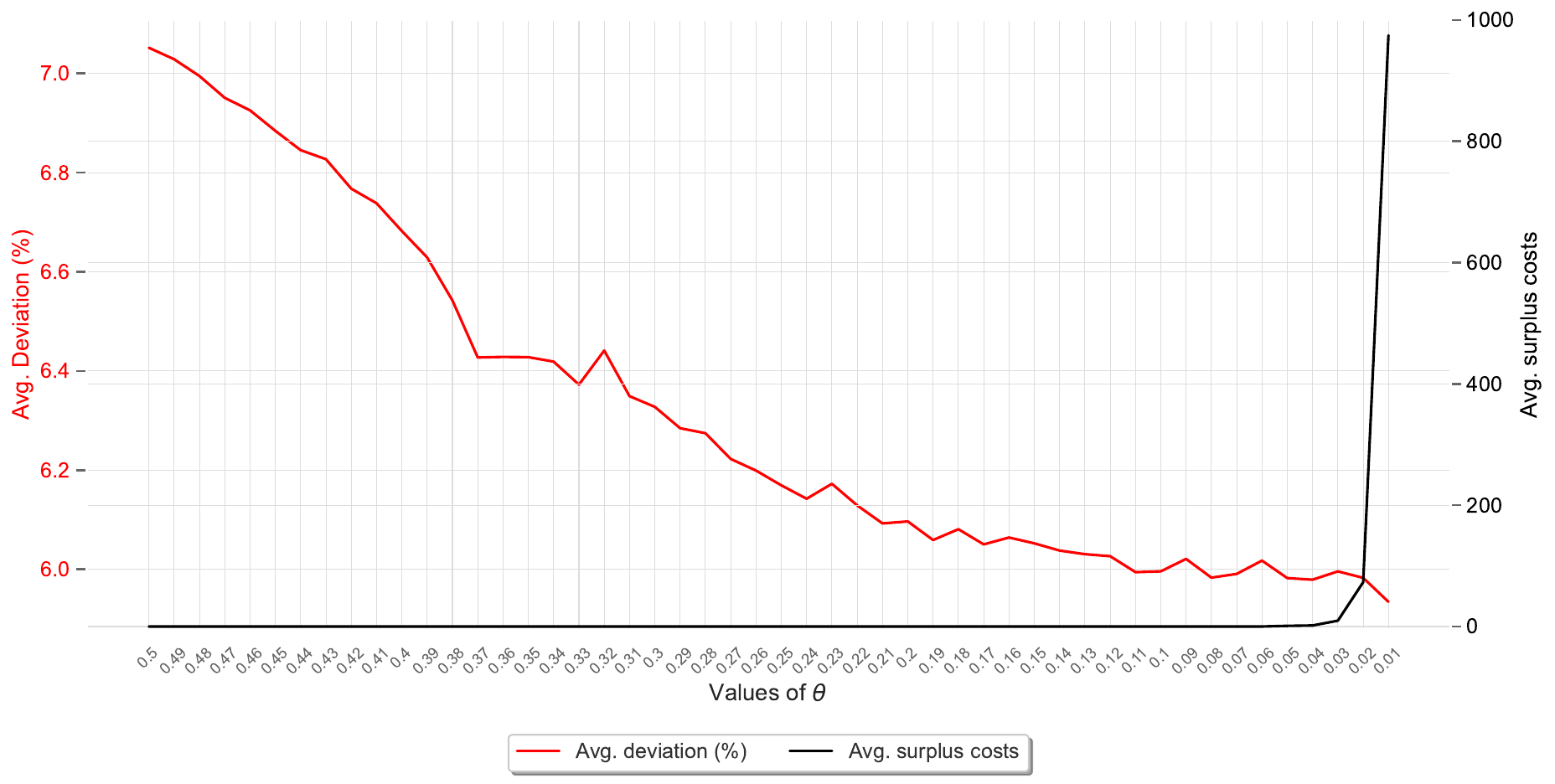}
  \caption{Long/short with round lots, various values of $\theta$.}
  \label{fig009}
\end{figure}

For both long only and long/short experiments, we vary $\theta = (0.50, 0.49, \dots, 0.01)$. Figures \ref{fig008} and \ref{fig009} display the long/short results with odd and round lots\footnote{The long only figures are shown in the appendix.}. In the figures, the left $y$-axis shows the average total deviation, calculated as 100 $\times$ the sum of all individual deviations for all assets and rebalances, divided by the number of rebalances. The right $y$-axis shows the average surplus in both fractional and borrowing costs, calculated as the sum of all surpluses for all assets in all rebalances divided by the number of rebalances.

With odd lots, the average deviation drops from around $0.44\%$ when $\theta = 0.5$ to around $0.39\%$ when $\theta = 0.01$. Meanwhile, the cost surplus is zero from $\theta = 0.50$ down to $\theta = 0.04$, sharply increasing when $\theta \leq 0.02$. We empirically observe the average deviation plateauing for $\theta < 0.2$. With round lots, the average deviation drops from around $7.1\%$ when $\theta = 0.5$ to around $6.0\%$ when $\theta = 0.01$. The average deviations are higher as round lots are much more restrictive, but they also drop more strongly with no obvious plateau. At the same time, costs surpluses increase rapidly for the smallest $\theta$ values.

In order to choose the ``sweet spot'' for $\theta$, we take into consideration that our main goal is minimising deviation, and that the further to the right on the chart, the smallest the average deviation. However we do not wish to distort the results by accepting large surpluses, although at least for the long/short strategy we might consider a minor (negligible) surplus to be acceptable. Hence we set, as an appropriate empirical choice, $\theta = 0.05$. The long only results shown in the appendix also support this choice.

\subsection{Case study: leverage}
\label{sec:marketNeutral}

Let $\beta_{\text{S\&P500}} = \frac{\sigma(r, b)}{\sigma(b, b)}$, where $r$ represents the out-of-sample returns of the investment strategy, $b$ the S\&P500 returns and $\sigma(x, y)$ the sample covariance between $x$ and $y$. Traditionally, one expects a long only strategy that only invests in S\&P500 stocks to have $\beta_{\text{S\&P500}}$ close to 1. In a market neutral strategy, on the other hand, we hope for $\beta_{\text{S\&P500}}$ to be close to 0. The potential advantage of this approach is reducing exposure to market risk.

A common way to create market neutral portfolios is by purchasing short positions in the index. Consider a portfolio composed of 50\% in long positions in S\&P500 stocks and 50\% in an unleveraged short position in S\&P500 futures. If we isolate the long part of the portfolio and find that it has $\beta_{\text{S\&P500}} = 1$, then we have a perfect market neutral strategy. The only problem is that such strategy is unlikely to generate any profits. We may adopt then a semi-active strategy by looking to outperform the index while enforcing market neutrality: with this objective we employ the same SSD approach from \cite{valle2017b} for choosing SSD-dominating ``market neutral'' portfolios. This may not yield exactly $\beta_{\text{S\&P500}} = 0$, but may be good enough for the purpose of the investment policy in effect.

When we account for leverage, we have further options in constructing market neutral strategies. If we adopt a $2\times$ leverage for the short position in S\&P500 futures, then we might invest $66.7\%$  in long and $33.3\%$ in short with the same effect. Likewise, with $3\times$ leverage we can adopt $75\%$ and $25\%$ proportions. A potential advantage of this approach is that the solver has more ``freedom'' in building SSD-dominating portfolios by having more capital available to invest in long positions.

In this section, we illustrate this approach. We adopt an odd lots policy and simulate the long only strategy from Section \ref{sec:caseStudy} without any derivatives. We restrict the out-of-sample period to be from $31^{\text{st}}$ December 2019 until $29^{\text{th}}$ December 2023. Let \texttt{SP500FUT} be the identifier for S\&P500 futures. We simulate market neutral strategies by fixing the short proportion in \texttt{SP500FUT} with $\upsilon^- = \upsilon^+ = [0.50, 0.33, 0.25, 0.20]$, and assigning $L_{\text{SP500FUT}} = [1, 2, 3, 4]$ as its leverage level.

\begin{figure}[!ht]
\centering
\includegraphics[width=0.8\textwidth]{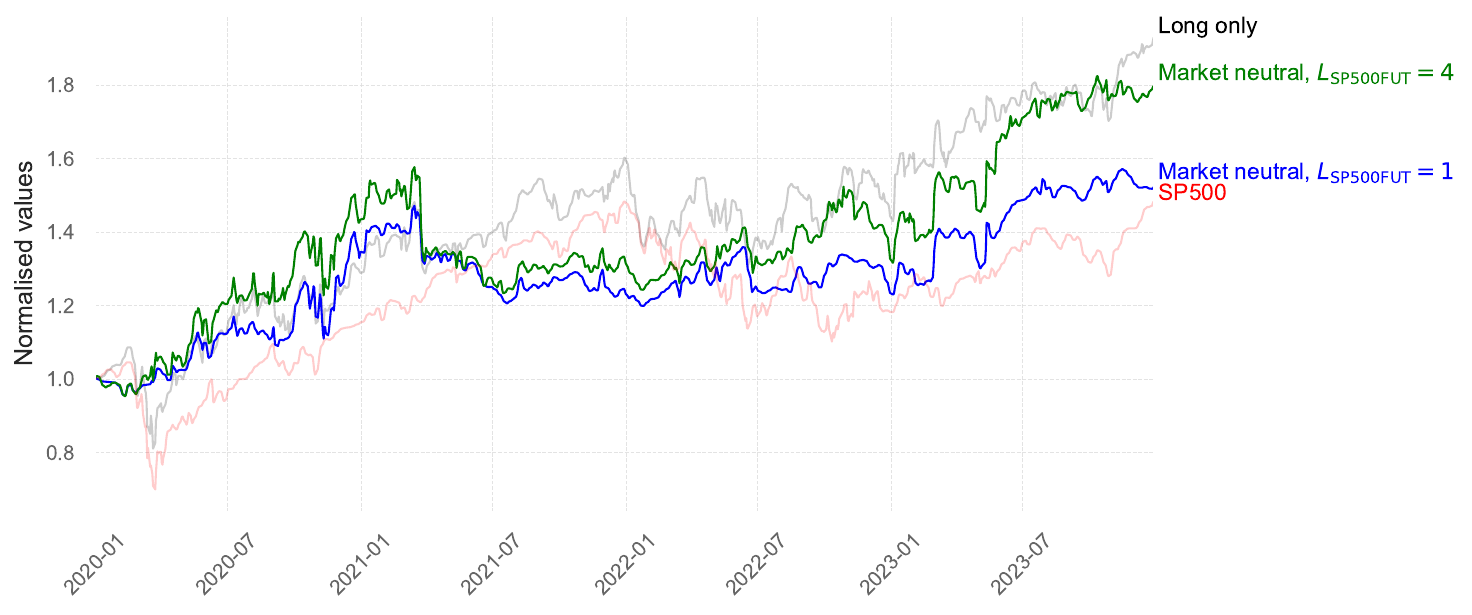}
  \caption{Out-of-sample cumulative performance, long only and long/short market neutral experiments.}
  \label{fig010}
\end{figure}

Results for these strategies are shown Figure \ref{fig010} and Table \ref{table11}. To prevent clutter, Figure \ref{fig010} displays only 3 of the 5 strategies, the lines are also smoothed with univariate splines.

\begin{table}[!ht]
\centering
{\scriptsize
\renewcommand{\tabcolsep}{1mm} 
\renewcommand{\arraystretch}{1.4} 
\begin{tabular}{|l|rrrrrrr|}
\hline
\multicolumn{1}{|c|}{Strategies} & \multicolumn{1}{c}{FV} & \multicolumn{1}{c}{CAGR} & \multicolumn{1}{c}{Vol} & \multicolumn{1}{c}{MDD} & \multicolumn{1}{c}{Sharpe} & \multicolumn{1}{c}{Sortino} & \multicolumn{1}{c|}{$\beta_{\text{S\&P500}}$}\\
\hline
Long only                                 &     1.93 &    17.90 &    24.45 &    26.06 &     0.70 &     1.00 &    0.721\\
Market neutral, $L_{\text{SP500FUT}} = 1$ &     1.52 &    11.11 &    18.17 &    18.74 &     0.55 &     0.78 &    0.114\\
Market neutral, $L_{\text{SP500FUT}} = 2$ &     1.71 &    14.33 &    20.26 &    19.23 &     0.65 &     0.95 &    0.118\\
Market neutral, $L_{\text{SP500FUT}} = 3$ &     1.74 &    14.88 &    21.18 &    22.66 &     0.65 &     0.96 &    0.100\\
Market neutral, $L_{\text{SP500FUT}} = 4$ &     1.80 &    15.79 &    21.42 &    22.35 &     0.69 &     1.01 &    0.093\\
SP500                                     &     1.48 &    10.24 &    23.01 &    33.92 &     0.44 &     0.61 & --      \\
\hline
\end{tabular}
}
\caption{Comparative out-of-sample statistics}
\label{table11}
\end{table}

Table \ref{table11} displays the same statistics as Table \ref{table6}, with $\beta_{\text{S\&P500}}$ as an additional column. Here the long only strategy outperformed the S\&P500, growing 93\% in the period, but also had a (high) $\beta_{\text{S\&P500}} = 0.721$. By adopting the unleveraged market neutral strategy, $\beta_{\text{S\&P500}}$ was reduced to $0.114$, but we also reduced the out-of-sample returns (up 52\%) by adopting a short position in S\&P500 (which increased 48\% in value during this period). By employing leverage, we are able to gradually improve returns (up to 80\% with $L_{\text{S\&P500}} = 4$) while maintaining similar levels of $\beta_{\text{S\&P500}}$. Note that, by adding leverage, we also observed a small increase in risk (Vol, MDD), not enough however to offset the increase in risk-adjusted returns (Sharpe, Sortino). Even with $4\times$ leverage, both drawdown and volatility were lower than the corresponding statistics for both S\&P500 and the long only strategy.

In summary, if used with care, leverage has the potential to improve risk-adjusted returns. This was a key factor when deciding to include futures contracts in our two-stage framework. In our view, this added flexibility opens up different research possibilities.

\section{Managerial insights}
\label{sec:managerial}

So far, we have considered several real-world features in our standardised two-stage framework (including those in the appendix). While we hope that this helps bridging the gap between theory and practice in portfolio selection, the list of features included is far from exhaustive. Different exchanges enforce different rules and financial instruments are subject to area-specific conditions. In this section we discuss practical aspects and features not yet considered, and which could result in future extensions of this framework.

In a practical sense, we operate under the assumption that live (non-delayed) prices $V_i$ are always available - as they are required by the second stage. Another implicit assumption we made is that, once a second stage decision is made, assets are to be traded at prices $V_i$. In cases where solving the second stage requires a few seconds (or more), real trading prices might substantially deviate from $V_i$, which in extreme cases could render the suggested portfolio infeasible in practice. To mitigate this issue, one could (i) trade at the end of the day, which is generally less volatile and/or (ii) impose a penalty at the price used in the second stage (if we are expecting an asset to be bought, increase its current price by a factor). If the second stage execution time is negligible, most likely these changes will be minor and random (sometimes upwards, sometimes downwards), with little observable effect. 

If prices deviate, special care must be taken in an automated trading environment: if the underlying system is designed to send trades at $V_i$, price deviations might cause trades to be cancelled. In this case a confirmation step must be in place to assess whether all trades were successfully executed. If not, then they may or may not be executed at a different price. Regardless, this information must be recovered and updated in order to minimise the deviation between the real and simulated investments.

Another important issue is liquidity. The second stage may suggest trades that are too illiquid when we compare their size to the corresponding average trading daily volume. There are three possible ways to tackle this: (i) by adding turnover constraints in the first stage, (ii) by limiting the size of trades that can be recommended in the second stage as a function of the investment size and volume or (iii) by adopting trading policies after the second stage. One example of policy commonly adopted (for infrequently rebalanced portfolios) is to split large trades into small ones, executed throughout multiple days. A possible extension of the two-stage approach is to make it multistage: Within the next days, suggest trades in multiple days such that not only the final portfolio matches the suggested as close as possible, but so do the temporary portfolios within the limits of liquid trades.

\cite{bertsimas1999} proposed a two-stage approach which uses relative variables in both stages. It was constructed to satisfy the requirements of a specific enterprise with specific goals, also since it does not use absolute variables some functions (such as costs and liquidity) have to be approximated rather than modelled with real values. However some considerations under that approach could be adopted into the second stage, such as minimising the number of trades or the deviation not only per individual asset, but also per asset class. Another possibility is to reformulate the objective function of the second stage in order to minimise the maximum deviation instead of the sum of deviations.

We have also not included dividends nor other cost functions. We have dealt with dividends so far at data level by assuming adjusted prices immediately prior to solving both stages. Regarding costs, our limited practice suggests that other cost functions such as fixed or piecewise convex or concave cost are either less common or can be eliminated through negotiation with brokers (although the latter is sometimes used to approximate liquidity). The second stage could however be extended to include these.

Finally, we have not considered the issue of taxes, which might have to be paid when closing certain positions. This is very region-specific; in certain situations we have observed that funds are exempt from paying taxes during operations, but have to pay them when clients request a withdrawal. We have also observed a situation where the government charges taxes periodically. We are yet to study how to incorporate these generically into the two-stage framework.

\section{Conclusions}
\label{sec:conclusions}

In this paper, we focus on the practical challenges of portfolio optimization, particularly the translation of optimal weights into actionable trades. We critically evaluate the existing scientific literature and identify several key issues that limit its practical application. To illustrate this, we present an empirical example using a single-stage approach from the literature to solve a basic portfolio problem (maximizing expected return). Our findings suggest that solvers often struggle to produce feasible or accurate solutions, highlighting the need for more robust methods.

With that in mind we propose a two-stage approach with the goal of building portfolios that can be directly implemented in real-world investment. Through extensive computational experiments, we show that our method is effective for reasonably diversified portfolios. While our two-stage approach is heuristic - since the second stage seeks to approximate the first stage solution - comparisons with an existing method show that, when it does not fail, the benefits of a single stage over a two-stage approach are somewhat marginal. Moreover, the two-stage method approaches exact solutions when the sums involved are large.

We believe this approach helps in reducing the gap between the theory of portfolio optimisation and the practice of investing. One particular application is in enabling fully automated investment. Moreover, by separating the problem in two, it is possible to add other realistic features without adding too much complexity to any model. We do so by considering, for the first time, portfolios composed of both futures contracts and equities together. We also propose the inclusion of borrowing costs in short positions.

Several research lines can be explored as future work, some of them have been mentioned in Section \ref{sec:managerial}. On top of that we also hope to expand the framework to support other asset classes, such as options and fixed-income instruments. We intend to explore combinatorial optimisation techniques and possibly alternative reformulations for improving the second stage computational performance in more extreme scenarios. 

Finally, in this paper we implicitly assumed that the portfolio decision framework is ``complete'' in the sense that a single agent is in charge of both choosing the optimal allocation and executing the purchases. Often, in investment funds, portfolios are optimised at a higher-level to either decide the allocation between asset classes or between financial intermediaries. In these cases, neither our two-stage nor the other existing formulations are directly applicable. As future work, it might be interesting to adapt the two-stage framework to define the capital to be allocated to each party in a decentralised scenario \citep{benita2019, benita2022}: the first stage considers intermediary portfolios as synthetic assets and the second stage, instead of considering lots and costs, could discount, for instance, fees charged by intermediaries or implicit costs of investment in different markets.

\section*{Conflicts of interest}
\noindent None declared.

\section*{Acknowledgements}
\noindent Cristiano Arbex Valle was funded by FAPEMIG grant APQ-01267-18.

\bibliographystyle{plainnat}
\bibliography{portfolioArXiv}

\end{document}